# Atomistic scale analysis of the carbonization process for C/H/O/N-based polymers with the ReaxFF reactive force field

Malgorzata Kowalik[1,‡], Chowdhury Ashraf[1,‡], Behzad Damirchi[1], Dooman Akbarian[1], Siavash Rajabpour[2] and Adri C. T. van Duin[1,*]

[1] Department of Mechanical Engineering, The Pennsylvania State University, University Park, PA 16802, United States

[2] Department of Chemical Engineering, The Pennsylvania State University, University park, PA, 16802, United States

**ABSTRACT**

During the carbonization process of raw polymer precursors, graphitic structures can evolve. The presence of these graphitic structures affects mechanical properties of the carbonized carbon fibers. To gain a better understanding of the chemistry behind the evolution of these structures, we performed atomistic scale simulations using the ReaxFF reactive force field. Three different polymers were considered as a precursor: idealized ladder PAN (polyacrylonitrile), a proposed oxidized PAN and PBO (poly(p-phenylene-2,6-benzobisoxazole)). We determined the underlying molecular details of polymers conversion into a carbon fiber structure. Since these are C/H/O/N-based polymers, we first developed an improved force field for C/H/O/N chemistry based on the Density Functional Theory (DFT) data with a particular focus on $N_2$ formation kinetics and its interactions with polymer-associated radicals formed during the carbonization process. Then, using this improved force field, we performed atomistic scale simulations of the initial stage of the carbonization process for the considered polymers. Based on our simulation data we determined the molecular pathways for the formation of low-molecular weight gas-species, all-carbon rings crucial for further graphitic structures evolution and possible alignment of the evolved all-carbon 6-membered rings clusters.

* Corresponding Author: Adri C. T. van Duin (acv13@engr.psu.edu)

‡ Equal contribution

# INTRODUCTION

Although carbon fibers have been used to manufacture the lighter and stronger materials since 1970[1], there is a constant need to lower production cost without compromising mechanical properties. Since at least 50% of the current cost of carbon fiber production is the cost of the commonly used polymer precursor, PAN (poly acrylic nitrile)[2], one possible option for lowering production cost of carbon fibers is to consider a different polymer as the precursor. A range of different laboratory engineered, as well as natural, polymers are being intensively investigated for this purpose, as an alternative route for carbon fiber production[2,3]. While there are multiple successful examples, such as pitch or lignin-based produced fibers[4], the constraints regarding required mechanical properties for the carbon fibers can still be challenging[5]. One of the main structural components responsible the mechanical properties of carbon fibers are graphitic structures that evolve during the carbonization process[3]. The evolution of these graphitic structures depends on the carbonization conditions (e.g. temperature and carbonization time) as well as the details of the molecular structures of the raw precursor fibers. Obtaining a deeper knowledge of the underlying molecular rearrangements that take place during carbonization process is an important step to understand evolution of graphitic structures. This knowledge can help us to eventually identify a possible candidate for a satisfactory polymer precursor.

Reactive atomistic simulations can be a very useful tool to gain information about chemistry behind molecular changes during the carbonization process and possible graphitic structures evolution. The ReaxFF reactive force field[6,7] has been successfully applied to a wide variety of systems and their properties[8], in particular for graphene mechanical properties investigations[9] or oxidation and pyrolysis of hydrocarbon fuels[10]. To date, simulations of the carbonization process using ReaxFF force field have been presented only for idealized ladder PAN[11]. These idealized ladder PAN molecules consist of C/H/N atoms only and the need for an improved force field, with parameters extended to C/O/H/N chemistry, has been suggested[11] so that more realistic, oxidized PAN molecules could also be considered in the reactive simulations. Here we present an improved ReaxFF force field parameter set, suitable for such C/H/O/N-based polymers simulations. With use of these new ReaxFF force field parameters, we investigated the carbonization process of PAN (idealized ladder and oxidized) and compared it with a possible alternative carbon fiber C/H/O/N-based polymer precursor: PBO (poly(p-phenylene-2,6-benzobisoxazole))[12,13]. PBO is considered to be an alternative carbon fiber precursor, since it was shown that an oxidation step, prior to the carbonization in the carbon fiber production, is not required for the PBO fibers, while such an oxidation step is necessary for PAN fiber polymers, increasing its production costs. PBO is a synthetic polymer with a rigid structure free of any potentially fuel-forming aliphatic -CH groups. The PBO fibers are known for their inherent superior thermooxidative stability, excellent tenacity, high resistance to solvent and easy processing properties[14]. While PBO is currently expensive, this wide range of useful properties ensures a constant rise in commercial applications of these fibers that will likely lead to a significant price reduction in the near future.

Based on the simulations of the carbonization process for these three different polymers (idealized ladder PAN, oxidized PAN and PBO, as shown in Fig. 1), we want to gain a deeper understanding of how the underlying molecular details of these polymers relate to the evolution of graphitic structures. All considered samples have the same thermal history and details of the carbonization simulations are presented later. First, we aimed to find out if the atomistic differences in the molecular structure of the PAN polymers, the idealized ladder PAN and oxidized PAN, can affect the nitrogen molecule production as well as all-carbon rings production. Then, in order to extend the question regarding possible correlations between underlying molecular details of precursor molecules and all-carbon ring production, we compared the data for the carbonization simulations for oxidized PAN and proposed alternative precursor, PBO. We also tested if the differences in the initial configurations, with polymers placed randomly in a box, vs aligned in one direction, could affect our results, for example by enhancing all-carbon rings evolution due to initial particular arrangement of nitrogen atoms. Further, based on the performed simulations, we proposed small molecules and all-carbon rings production pathways and assessed possible 6-membered rings alignment for all considered polymers.

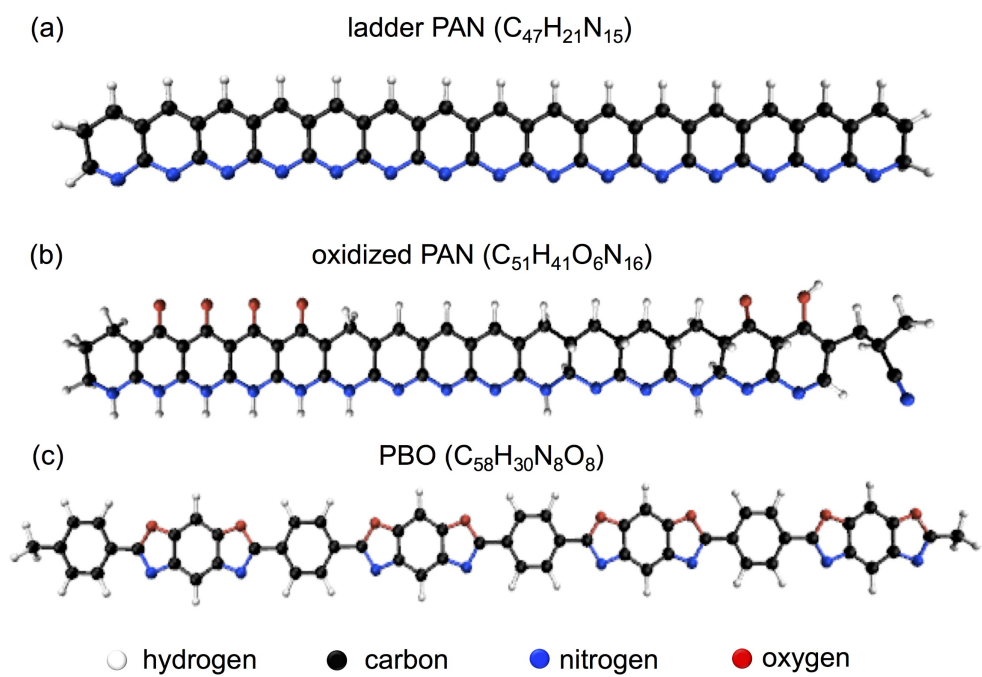

Figure 1: **The molecular structures of the considered polymers.** The molecular structures of: (a) ladder PAN (idealized structure with no oxygen atoms), (b) proposed oxidized PAN (polyacrylonitrile)[32] and (c) PBO (poly(p-phenylene-2,6-benzobisoxazole)) with 4 repeat units.

## METHODS

## SIMULATIONS DETAILS

**ReaxFF Reactive Force Field**

The ReaxFF reactive force field based method can simulate bond formation and bond breaking as they occur during molecular dynamics simulations, thus enabling simulation of chemically reactive systems. ReaxFF uses a bond order-bond distance[15,16] relation in conjunction with the bond order-bond energy relation, which enables it to properly simulate the smooth formation and dissociation of bonds. All the connectivity dependent terms like bond, angle and torsion are made bond order dependent so that their contribution will diminish if the bond breaks. However, non-bonded interactions like van der Waals and Coulomb are calculated between every pair of atoms irrespective of their connectivity. Though the non-bonded interactions are not bond order dependent, they are highly dependent on the distance of the atom pairs, so these contributions need to be updated at each iteration. ReaxFF calculates atomic charges using a geometry-dependent charge calculation scheme and uses Electronegativity Equalization Method (EEM)[17] for this purpose. Additionally, to eliminate discontinuities in the non-bonded interaction energies and to reduce the range of the Coulomb-interactions, a seventh order Taper function is employed,[6] which ensures that all non-bonded terms, together with their first, second, and third derivatives, go to zero at the non-bonded cutoff distance, which is typically picked to be 10 Ångstrom. In short, ReaxFF uses the following equation to find the energy and then the force on each atom:

$$E_{\text{system}} = E_{\text{bond}} + E_{\text{over}} + E_{\text{under}} + E_{lp} + E_{val} + E_{tor} + E_{\text{vdWaals}} + E_{\text{coulomb}} + E_{trip}, \quad (1)$$

where, $E_{bond}$, $E_{over}$, $E_{under}$, $E_{lp}$, $E_{val}$, $E_{tor}$, $E_{vdWaals}$, $E_{coulomb}$, $E_{trip}$ represent bond energy, over-coordination energy penalty, under-coordination stability, lone pair energy, valence angle energy, torsion angle energy, van der Waals energy, Coulomb and triple bond stabilization energy, respectively. For a more detailed description, readers are referred to the original ReaxFF papers.[6,7] Since its development, ReaxFF molecular dynamics simulations have been used to study varieties of processes including combustion,[7,10,18–20] battery materials,[21,22] fuel cell polymers,[23] polymer cross-linking and mechanical properties,[24] metal-oxide interfaces,[25] catalysts,[26] 2D materials[27] and many more. A recent review[8] of the ReaxFF method and its applications summarizes the capability and future directions of the method.

**Parameterization of ReaxFF CHON force field**

The quality of a molecular dynamics simulation depends on the accuracy of the force field parameters, therefore, these parameters needed to be trained against available experimental or quantum mechanical (QM) based density functional theory (DFT) data. Our system of interest (carbon fiber precursors) contains C, H, O and N atom types, therefor we started our force field training with recently developed CHO-2016 force field by Ashraf et al.[28] and added the N atom type in the force field. Since CHO-2016 is trained for C/H/O chemistry, especially for chemistry of small molecules, we only needed to train the bond/angle/dihedral parameters that involve any N atoms. Here, it is important to mention that the carbon parameters of CHO-2016 parameters are taken from Srinivasan et al.[9], which were basically developed for obtaining better mechanical properties for graphene. Thus, these parameters are very much relevant to this project as we would like to estimate the mechanical strengths of the generated carbon fibers.

Previously, ReaxFF simulations of polyacrylonitrile (PAN) as carbon fiber precursor were performed by Saha et al.[29] in 2012, where they used a force field developed by Kamat et al.[30]. That force field was basically built upon the CHO-2008 parameters by Chenoweth et al.[7] where atom type N was added to simulate carbon fiber precursors. We will refer this force field by CHON-2010 as this force field was first published on 2010. Though the results obtained by Saha et al.[29] were encouraging, we had a three-fold motivation for developing a new set of ReaxFF parameters for such systems:

1. The carbon parameters of CHON-2010 are not reasonable enough for estimating graphene mechanical properties.[9]

2. Our simulations indicated that $N_2$ molecules in CHON-2010 are less stable and quickly react with carbon radical site, therefore, no $N_2$ molecule can be produced during the simulation without actively removing them from the simulations box.

3. Besides PAN, we would also like to simulate oxidized PAN and PBO molecules, which has oxygen atoms. Therefore, developing a new force field containing all four atom types (C/H/O/N) will give us more flexibility for that purpose.

Taking all the above three points under consideration, we started our force field training taking CHO-2016 parameters developed by Ashraf et al.[28] as a basis and started adding N parameters in the force field. This newly developed force field will be referred to as CHON-2019 from hereon. In CHON-2019, we trained different possible bonds, angles and torsion parameters that involve at least one nitrogen atom, and several key training results can be found in the *supporting information*. During training we found that the valence angle modification proposed by Ashraf et al.[28] does not agree well when the angle consists of nitrogen atom(s), especially since it makes HCN an angular molecule while it is basically a linear one. Therefore, we removed this correction for any angle with nitrogen atom(s). Additionally, in order to make $N_2$ molecule more stable i.e. less reactive in CHON-2019, we implemented the triple bond stabilization term ($E_{trip}$, Equation 2), which is already in place for other stable triple bonds such as those in CO molecules. In this force field, this stabilization energy is only active for CO and $N_2$ molecules, not for other stable triple bonds such as HCN. An additional parameter in the general section of the force field (vpar(40)) is introduced to switch on/off this stabilization energy contribution. If vpar(40) = 0, then the stabilization only applies to CO molecule. If vpar(40) = 1, then all the possible triple bonds receive this contribution, while if vpar(40) = 2, only CO and $N_2$ molecules are affected. These changes are implemented both in stand-alone ReaxFF source code and in ADF modeling suite[31] for parallel implementation:

$$E_{trip} = p_{trip1} \exp\left[-p_{trip2}(BO_{ij} 2.5)^2\right] \cdot \frac{\exp\left[-p_{trip4} \cdot \left(\sum_{k=1}^{neighbors(i)} BO_{ik} - BO_{ij}\right)\right] + \exp\left[-p_{trip4} \cdot \left(\sum_{k=1}^{neighbors(j)} BO_{jk} - BO_{ij}\right)\right]}{1 + 25 \exp[p_{trip3}(\Delta_i + \Delta_j)]}$$

(2)

Fig. 2 shows some significant improvements of CHON-2019 over the CHON-2010 force field. Here, we will only discuss the improvement of ReaxFF force field related to PAN chemistry. Fig. 2(a) shows the comparison between DFT and ReaxFF energies of the reaction $CH_3 + N_2 \rightarrow CH_3N_2$. This reaction is endothermic DFT, therefore $CH_3N_2$ is a less stable molecule making $N_2$ more stable in the system. However, CHON-2010 makes this reaction exothermic by almost 55 kcal/mol. Therefore, whenever an $N_2$ molecule is generated in CHON-2010, it prefers to react with an under-coordinated C atom in the system, which prevents $N_2$ molecule build up in the system. This is non-physical and introducing the triple bond stabilization term in CHON-2019 makes this reaction endothermic again, comparable with DFT results. In addition to that, we compared the relative energies of formation of ladder PAN structure from linear PAN, shown in Fig. 2(b). Again, CHON-2019 parameters give much better agreement with DFT than the previous version. Furthermore, we compared the $N_2$ removal reaction from two consecutive heterocyclic six member rings of ladder PAN between ReaxFF and DFT, this reaction mechanism of forming to neighboring five member carbon rings were reported by Saha et al.[29] The comparison between DFT and both the ReaxFF force fields are shown in Fig. 2(c) indicating that CHON-2019 energetics are in good agreement with DFT. Lastly, we trained the force field against a Diels-Alder type reaction where two HCN and one $C_2H_2$ molecules react to generate one heterocyclic six-member ring (Fig. 2(d)), where again CHON-2019 gives much better agreement than CHON-2010 with the DFT numbers. A detailed comparison between DFT, CHON-2010 and CHON-2019 energetics can be found in the supporting information section.

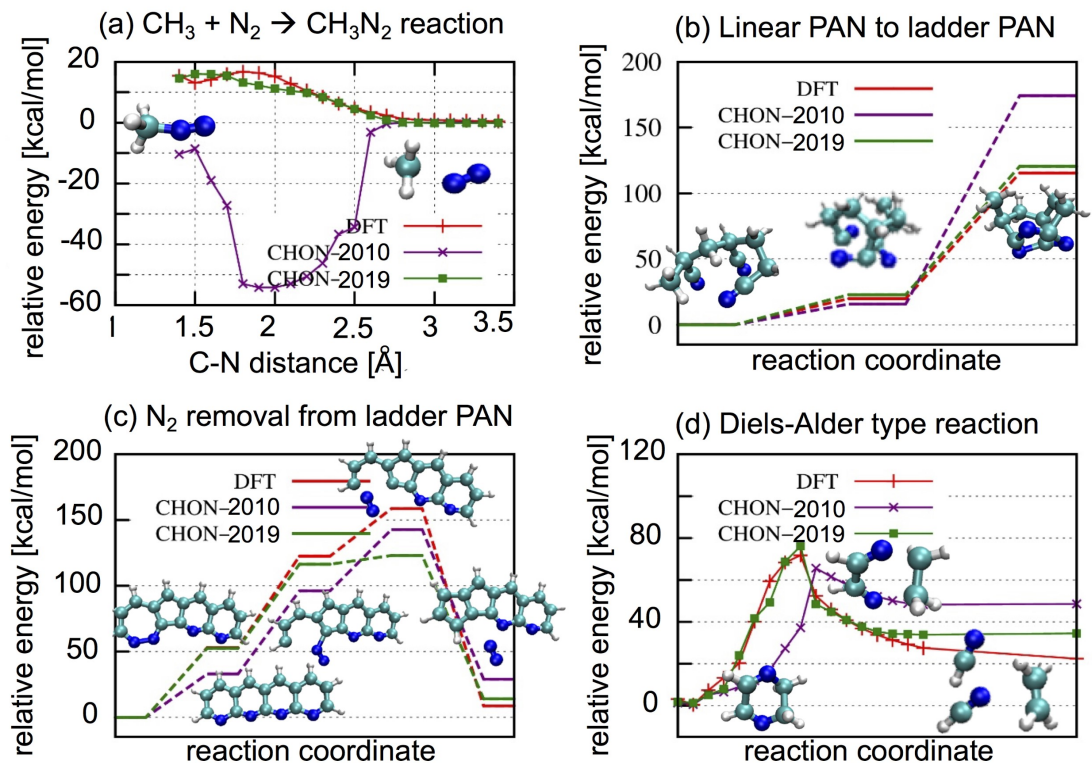

Figure 2: **Comparison between DFT and ReaxFF for PAN (polyacrylonitrile) related chemistry**. (a) Reaction energy for $CH_3 + N_2 \rightarrow CH_3N_2$ reaction. CHON-2019 force field predicts this reaction to be endothermic, same as DFT, while CHON-2010 makes it very exothermic. (b) Formation of ladder PAN from linear PAN, (c) Removal of N2 molecule from ladder PAN structure and formation of two neighboring five- member carbon rings. (d) The Diels-Alder type reaction involving HCN and C2H2 molecules. The DFT numbers are calculated using B3LYP hybrid functional and 6-311G** basis set. The cyan, blue and white spheres represent C, N and H atoms respectively.

**ReaxFF Simulations**

The molecular structures of the considered polymers are presented in Fig. 1. The PAN molecules were chosen to have 15 repeat units and oxidized PAN was built based on a proposed molecular model obtained from experimental data[32]. The PBO molecule has 4 repeat units, so the total number of carbon atoms in each considered polymer was comparable. Each of these polymers were initially energy minimized and then 16 polymers of the same type were placed in the simulation box, either randomly or aligned (Fig. 3 (a)), leading to 6 systems for consideration. All simulations, unless indicated differently, were performed using NVT ensemble, with constant number of molecules (N), volume (V) and temperature (T). We used the Berendsen[33] thermostat with a temperature coupling constant 100 fs. For all considered systems periodic boundary conditions were applied, the time step was chosen to be 0.25 fs, and a bond order cutoff for molecule identification of 0.3 was used. Initially, all simulations boxes were deformed, so the densities for all considered systems were 1.6g/cm$^3$, the approximate experimental density for PAN and PBO[12]. All these systems were then equilibrated for 100 ps at temperature of 300 K using NVT ensemble. The last three configurations, saved every 10 ps, for all systems, were chosen as an initial configuration, so we could have three different samples for each system. Then all samples were heated with rate of 10K/ps up to 2800K and then NVT simulations were performed for 900 ps. All presented simulations were performed with ADF simulation software[31].

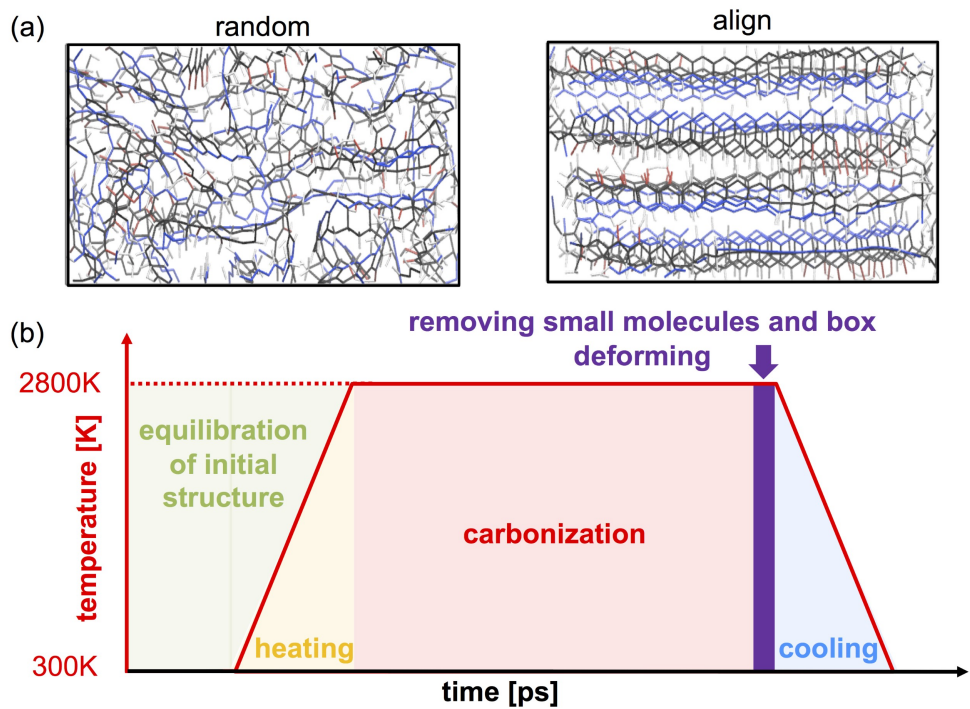

Figure 3: **The initial configurations and simulation scenario**. (a) The snapshots of the initial configurations for oxidized PAN with all molecules placed in the simulation box randomly or aligned along x-direction after 100 ps of equilibration simulations at NVT ensemble with T = 300K. For clarity, only bonds between the given type of atoms are shown, as black, white, red and blue, for C, H, O and N atoms, respectively. (b) Schematic representation of thermal history for all considered samples.

**RESULTS AND DISCUSSION**

As mentioned earlier, only the reactive simulations for idealized ladder PAN have been previously reported[29,34,35]. Equipped with the CHON-2019 ReaxFF parameters set we can determine if the structural differences of the proposed oxidized PAN and idealized ladder PAN see Fig 1 (a-b), can affect the simulation results. Particularly, we can test if these differences in the atomic structures of PAN precursor polymers can affect nitrogen molecules and all-carbon rings production. An average number and standard deviations of the nitrogen molecules produced in the case of the carbonization simulations for oxidized and ladder PAN are compared in Fig. 4 (a) for three samples per each initial configurations of the polymers, random or aligned. Due to the differences in the molecular structure of the considered polymer chains, with 16 nitrogen atoms for the oxidized PAN versus 15 nitrogen atoms for ladder PAN, we did plot a percentage of the produced $N_2$ molecules with respect to maximum possible number of these molecules produced for each of the systems. Based on the shape of the nitrogen molecules production curves, it is clear that we do have a systematically increasing production of these molecules. Once an $N_2$ molecule is produced, it does not react back with the rest of the molecules in the system. This inert characteristic of the nitrogen molecules is expected, due to the extremely strong triple bond that exists in the nitrogen molecule. However, the simulations of the ladder PAN system with use of the CHON-2010 ReaxFF force field show consistent $N_2$ production only during active removal of these molecules from the simulations box, as applied by Saha et al.[29]. We can also say that the number of the $N_2$ molecules produced in the case of oxidized PAN is higher for any of the considered initial configurations, aligned or random, compared to the idealized ladder PAN.

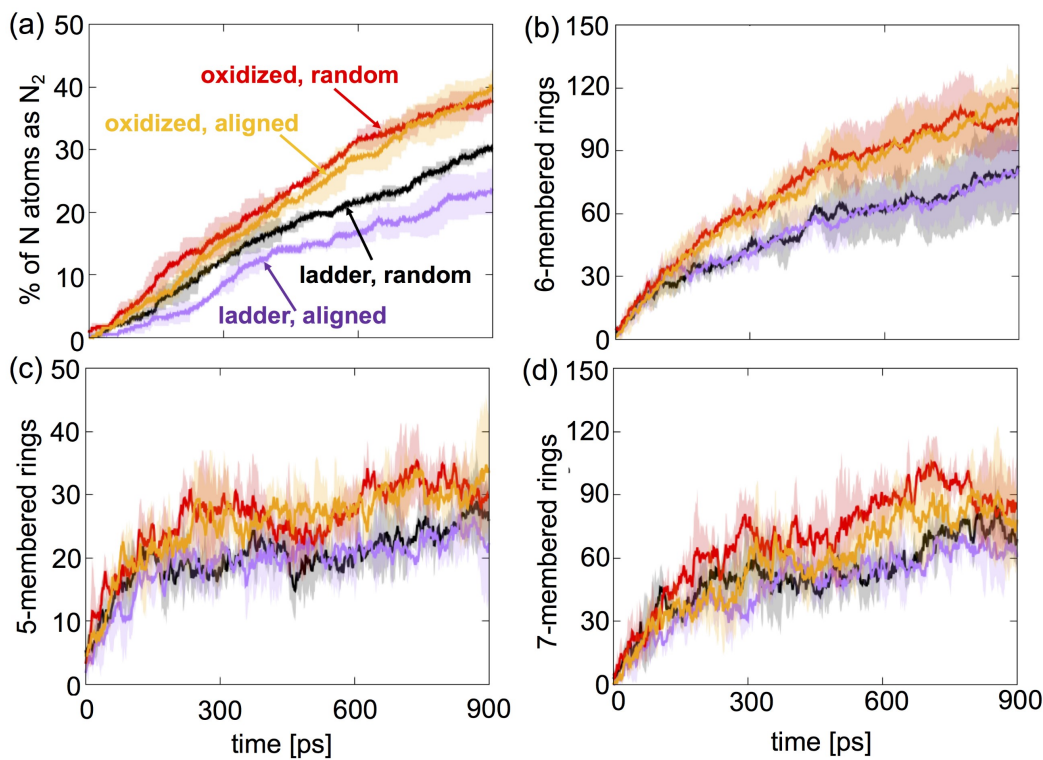

Figure 4: **The nitrogen molecules and all-carbon rings productions.** (a) A comparison of the nitrogen molecules productions for ladder PAN system and proposed oxidized PAN in the case of the all polymers initially placed randomly in the simulation box or aligned (see Fig. 3 (a)). The percentage is calculated with respect to maximum possible $N_2$ molecules that could be created for each system, based on the molecular structure of the considered polymers. (b)-(d) A comparison of 5/6/7-membered rings production for the same samples: ladder/oxidized PAN with random/align initial configurations. All data are obtain from NVT simulations at T= 2800K, after initial equilibration at 300K and heating the systems with constant heating rate 10K/ps. The presented curves represent the averaged values from 3 different samples for each system, where shadow areas represent the corresponding standard deviations.

In the case of the oxidized PAN, the initial alignment does not affect nitrogen molecule production. This is not true for ladder PAN, where for the initially aligned chains the $N_2$ production is the smallest. Since all simulations were performed in the same conditions, the origin of these discrepancies likely comes from the structural diversity of the polymer chains. The main differences in the molecular structures of ladder and oxidized PAN is the presence of the five carbonyls, one hydroxyl and 8 imide groups for the oxidized PAN, whereas in the case of ladder PAN there are no oxygen-containing groups and none of 15 nitrogen atoms are hydrogenated. For the oxidized PAN, the initial chain alignment (where chains are aligned such that nitrogen atoms from adjacent chains are the closest neighbors) is apparently not required for $N_2$ production, since for all samples with the polymers initially aligned or placed randomly in the box, we have a comparable number of nitrogen molecules produced. On the other hand, for ladder PAN we do have less $N_2$ molecules produced in the case of the polymer chains that are initially aligned, and this initial alignment makes the $N_2$ production even smaller. Based on these observations, we can conclude that the initial proximity of the nitrogen atoms is not required for the nitrogen molecules to be produced. As mentioned earlier regarding the structural differences between ladder and oxidized PAN, we can also speculate that the presence of imide groups can be helpful for this production. The evidence to support this speculation will be presented later based on the reaction pathways observed for the production of the gas molecules. Our carbonization simulations were performed at relatively high temperature, 2800K, compared to experimental temperatures range of 1800-2300K[36], so we can observe reasonable amounts of reactions at relatively short time scale (less than 1 ns). This might cause some discrepancies between the simulations and experimental data, like the temperature when $N_2$ molecules start to be produced. Even so, the experimental data for PAN precursor shows that nitrogen production is quite characteristic for temperatures > 800K[36], whereas the observed crystallinity for PAN fibers is not reported above 600K[37]. These experimental observations reflect our simulations data, which shows the comparable nitrogen molecules productions for oxidized PAN are independent of initial polymer alignment.

We also compared all-carbon rings production for ladder and oxidized PAN, as seen in Fig. 4 (b-d), for 5-/6-/7-membered rings, respectively. While 5- and 7-membered ring productions are only a little smaller for ladder PAN compared to oxidized PAN for all considered samples, the 6-membered rings production is significantly smaller for any of the ladder PAN samples compared to the oxidized PAN. This indicates that there is no simple relation between the all-carbon rings and nitrogen molecules productions, since we do not observe the same trend for these productions for ladder versus oxidized PAN samples. Since, the 6-membered rings are the basic building blocks for graphitic structure, the more 6-membered rings are in the system, the better the chance for graphitic structure to evolve. In Fig. 4 b) we see that the number of 6-membered rings is growing for all samples, whereas the number of 5-membered, Fig 4 (c), and 7-membered rings, Fig. 4 (d), starts to fluctuate around the constant values after around 500 ps of the carbonization simulations. We can compare the number of 6-membered rings produced in our simulations with the number of the sp2 carbons that we know from experimental data. For the carbonized fibers at least 90% of the sp2 carbons is reported[3]. The dominant and increasing number of 6-membered rings indicates the correct trend observed in our simulations. After 900 ps this number was approximately 60% of all rings in the system, for all our simulations, indicating an early stage of the carbonization process.

Furthermore, we can compare all-carbon ring production for oxidized PAN and the proposed alternative carbon fiber precursor, PBO. Based on this comparison, we can say that there are on average more all-carbon rings produced in the case of any PBO systems relative to the oxidized PAN ones. However, the percentage of 5-/6-/7-membered rings for all PBO samples is comparable to the same percentages for all oxidized PAN systems, that is approximately 20% of 5-membered rings, 60% of 6-membered rings and 20% of 7-membered rings. Moreover, in the case of PBO samples, we can notice a delay in the 6-membered ring productions for the polymers initially aligned, which appears to be correlated with the delay in the 5-membered ring production, see Fig. 5 (a-b). Based on this observation, we can suspect that 5-membered rings can often be the origins of the 6-membered rings. This observation was also proposed earlier, based on the ladder PAN simulations[29] and will be further discussed in the following material.

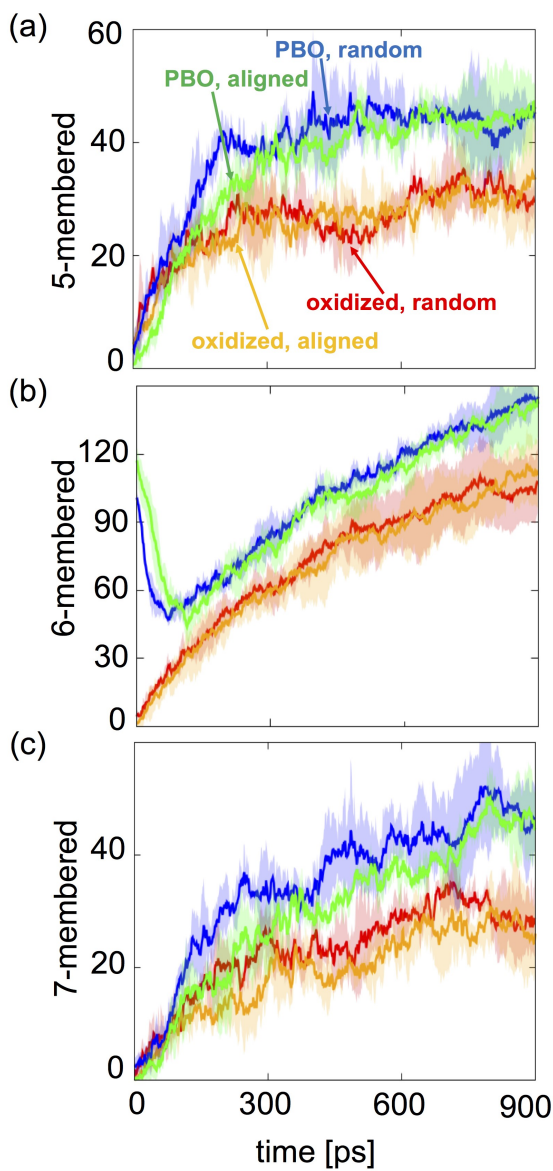

Figure 5: **A comparison of all-carbon rings productions for oxidized PAN and PBO.** A comparison of 5-/6-/7-membered rings production for the oxidized PAN and PBO with random/align initial configurations. All data are obtain for NVT simulations at T= 2800K, after initial equilibration at 300K and heating the systems with constant heating rate 10K/ps. The presented curves represent the averaged values from 3 different samples for each system, where shadow areas represent the corresponding standard deviations.

As is generally accepted, during the carbonization process small molecules dissociate from polymer chains and are released as gases. Based on the reactive simulations, we can make predictions regarding possible released gases and understand the origins of the differences in these small molecule productions based on the atomistic diversities of the considered polymer precursors. The comparison of the $H_2$, $H_2O$, CO and $CO_2$ production for the oxidized PAN and PBO is presented in Fig. 6. An overall relative ratio of these small molecules produced during the heat treatment observed in our simulations and the one reported based on the experimental data are in the reasonable agreement[38], with higher CO and $CO_2$ production observed for PBO and higher H$_2$O production observed for oxidized PAN samples. Based on the data plotted in Fig. 6, we can say that there are no significant differences in the considered small molecule productions for either oxidized PAN or PBO, based on the initial polymer configurations, random versus aligned, except for CO production for PBO. A delay in the production of carbon monoxide for the PBO samples with polymers initially aligned is observed. This delay can be correlated with the delay in the 5-membered rings production and 6-membered rings, that are part of the PBO polymer backbone opening (see Fig. 5 (a-b)) observed for these systems, compared to the ones with polymers initially randomly placed in the box. This suggests that a release of the CO molecule often result in 5-membered ring production and 6-membered rings opening. Overall, we observe higher H$_2$ and H$_2$O production for oxidized PAN compare to PBO, whereas in the case of CO and CO$_2$ production, the trend is reversed, with more carbon monoxide or dioxide produced for any PBO system compared to oxidized PAN. This observation confirms a role of the oxazole groups present in the PBO chains that are a source of the carbon monoxide and carbon dioxide molecules and are a starting point for all-carbon ring production. On the other hand, the hydrogen-based molecules, like $H_2$ and $H_2O$, appear to play this role in the case of the oxidized PAN systems. Based on the data reported in literature[39] we know that for oxidized PAN samples, and also ammonia and hydrogen cyanide production should be observed. However, we know also, that in the case of the elevated pressure and temperature, these molecules might be not stable. Dattelbaum et al.[40] report that for shockwave compression of ammonia gas they do observe conversion of the ammonia into $N_2$ and $H_2$ molecules. To test the possibility of production of HCN and $NH_3$ molecules without letting them react, we performed a simulation for the oxidized PAN sample while removing these CHN and $NH_3$ molecules from the simulation box every 2000 time steps and compared the data from this simulation with the data for the sample where no small molecules were removed, as seen in Fig. 7. We do observe a significant reduction in the hydrogen and nitrogen molecules productions for the system when ammonia and hydrogen cyanide molecules are systematically removed from the simulation box, as seen in Fig. 7 (a) and (c). Also, we observe a systematic production of these molecules, NH$_3$ and HCN as shown in Fig. 7 (b) and (d), in the case of the simulations were these molecules are removed during a course of the simulations. Moreover, if we compare the all-carbon rings production for these two cases, without and with systematic removal of the small molecules from the simulations box, see Fig. 8, we note that this removal affects mostly the 6-membered rings production, which is significantly reduced compared to the closed system simulations. These differences can be correlated with the experimental observation of the variations in the cross-sectional structure of the carbonized fibers[3], with the core part better represented by the simulations of the closed system, where no molecules are removed from the simulation box, and the skin part of the carbon fiber is better represented by the open system simulations.

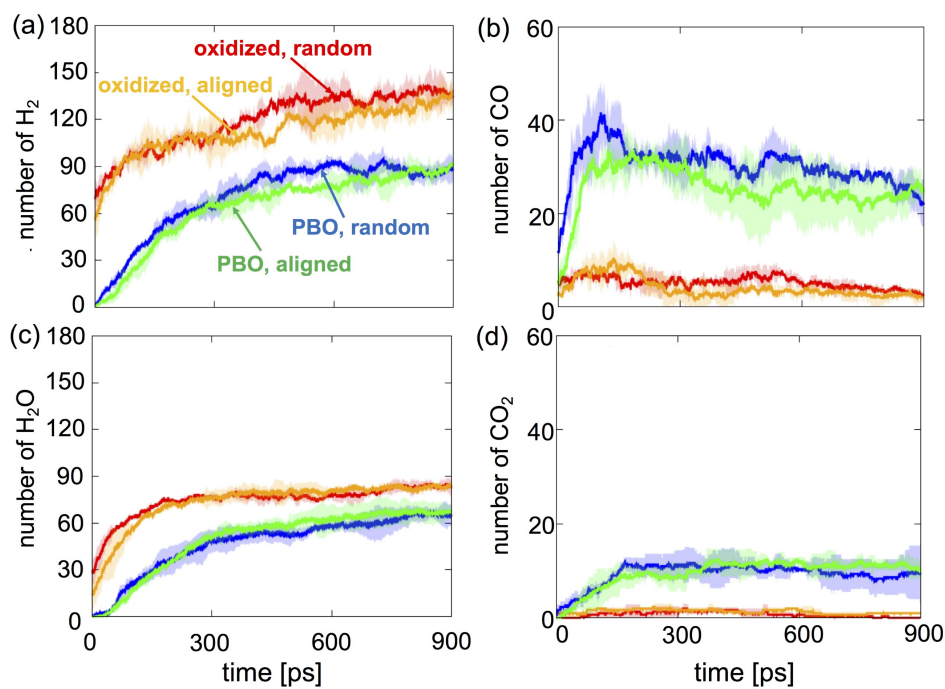

Figure 6: **The C/H/O molecules production. A comparison of C/H/O-atoms containing molecules production for oxidized PAN and alternative carbon fiber precursor:** PBO observed during the carbonization simulations at 2800K. The production of (a) hydrogen ($H_2$) molecules, (b) carbon monoxide (CO), (c) water ($H_2O$) and (d) carbon dioxide ($CO_2$). All data are obtained for NVT simulations at T= 2800K, after initial equilibration at 300K and heating the systems with constant heating rate 10K/ps. The presented curves represent the averaged values from 3 different samples for each system, where shadow areas represent the corresponding standard deviations.

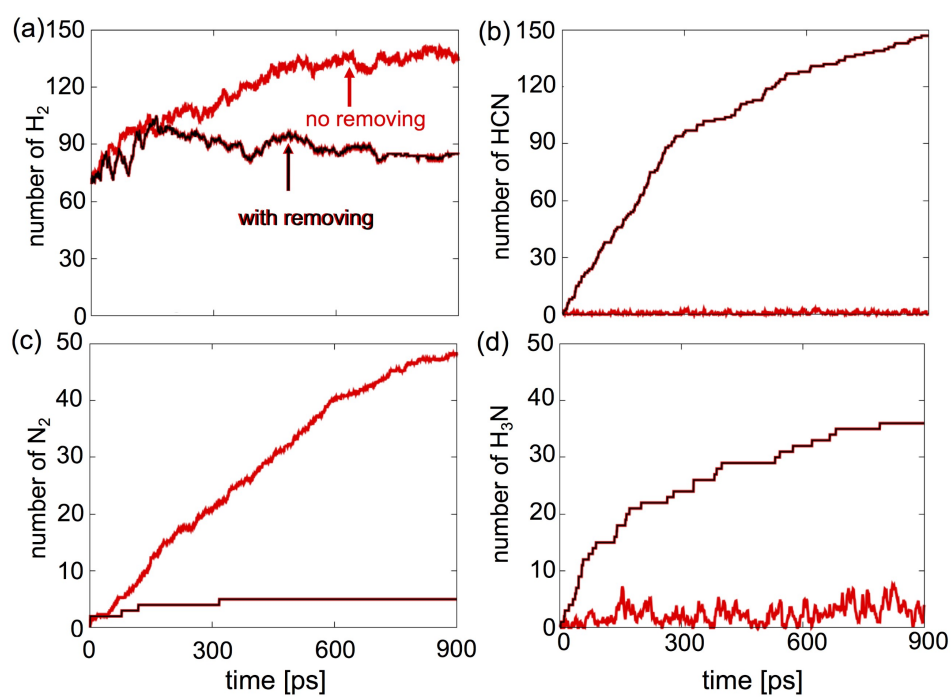

Figure 7: **The gas production for system with and without removing ammonia and hydrogen cyanide**. The comparison of the production of the (a) hydrogen (H$_2$) , (b) hydrogen cyanide (HCN), (c) nitrogen (N$_2$) and (d) ammonia (NH$_3$) molecules for the simulations of the same system performed at two conditions: where no molecules were removed (red curve) and when NH$_3$ an HCN molecules were removed every 0.5 ps during the course of the simulations (black-red line).

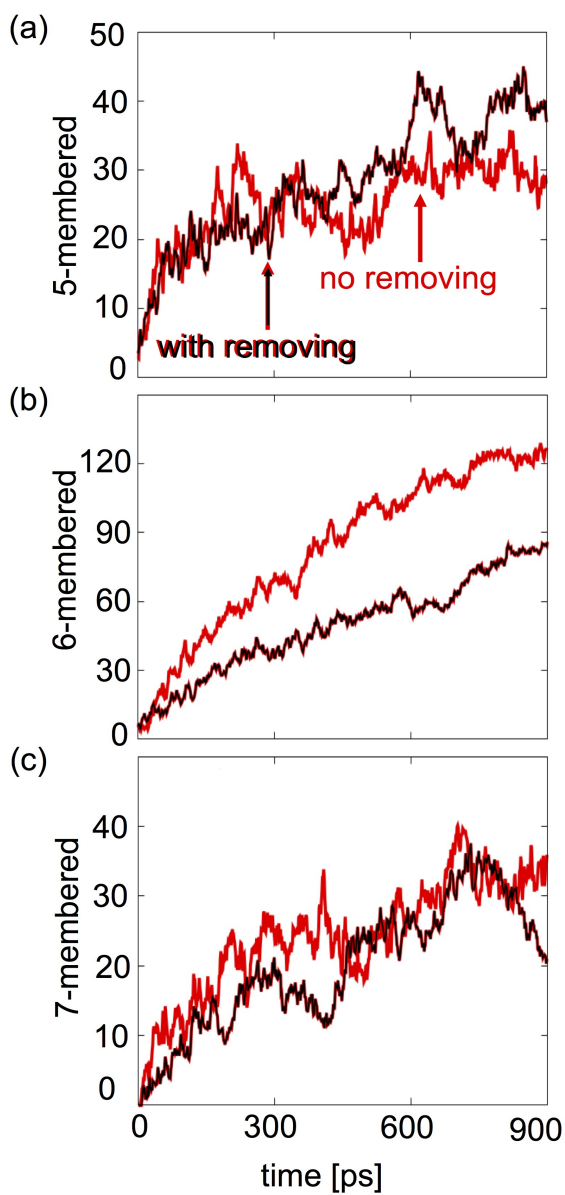

Figure 8: **The all-carbon rings production for system with and without removing ammonia and hydrogen cyanide**. The comparison of (a) 5-membered, (b) 6-membered (c) 7-membered rings production for the simulations of the same system performed at two conditions: where no molecules were removed (red curve) and when NH3 an CHN molecules were removed every 0.5 ps during the course of the simulations (black-red line).

A closer look into a simulation trajectory allows us to extract the possible pathways for the considered molecule production. While some scenarios were proposed earlier based on the experimental[36] or simulations data[29], we observe in our reactive simulations scenarios not reported previously. In Fig. 9 (a-b) the snapshots presenting a formation of the hydrogen molecules are presented. For each snapshot a frame number is indicated in the left corner. These frames were stored every 6.25 fs. The production of these molecules already starts during the heating stage of the simulations and the same pathways are observed for ladder as well as oxidized PAN. Whenever two hydrogen atoms from –NH group are close enough, a hydrogen molecule is produced. However, if the same groups, -NH, survive and meet at higher temperature (during the carbonization simulations at temperature 2800K) a bond between nitrogen atoms can be created and eventually nitrogen molecules are released, Fig. 9 (c-d). The remaining hydrogen atoms could form hydrogen molecule, as shown for ladder PAN, Fig. 9 (c), or could stay in the surrounding environment, as shown for oxidized PAN, Fig. 9 (d). Based on this knowledge, regarding possible nitrogen molecules production, we can explain our previous observations that there is reduced $N_2$ production for the ladder PAN compared to the oxidized PAN. For oxidized PAN systems there are –NH groups present along the polymer backbone, whereas in the case of ladder PAN none of the nitrogen is initially hydrogenated, see Fig. 1 (a-b). This explains why we do not observe a significant difference in $N_2$ production for oxidized PAN with chains initially aligned or not, as well as the reduced nitrogen molecule production for ladder PAN samples compare to the oxidized ones. For ladder PAN with polymers initially aligned in the box, it can be more difficult for nitrogen atoms to be hydrogenated, so we observe even more reduced $N_2$ production for systems with initial alignment for ladder PAN, see Fig. 4 (a). In the case of water molecules production, a proximity of –OH and –NH groups results in a release of the $H_2O$ molecule, Fig. 10 a-b), for both, oxidized PAN as well as PBO. However, in the case of the PBO systems, much more oxygen atoms contribute in a carbon monoxide release than water, as can be expected, since for CO to be released only C-O bond in the oxazole group needs to be broken, whereas for $H_2O$ to be released the nearby presence of –NH group is also necessary. In the case of oxidized PAN two C-C bonds need to be broken for one carbon monoxide molecule to be released, therefore we observe smaller carbon monoxide production for oxidized PAN compare to PBO.

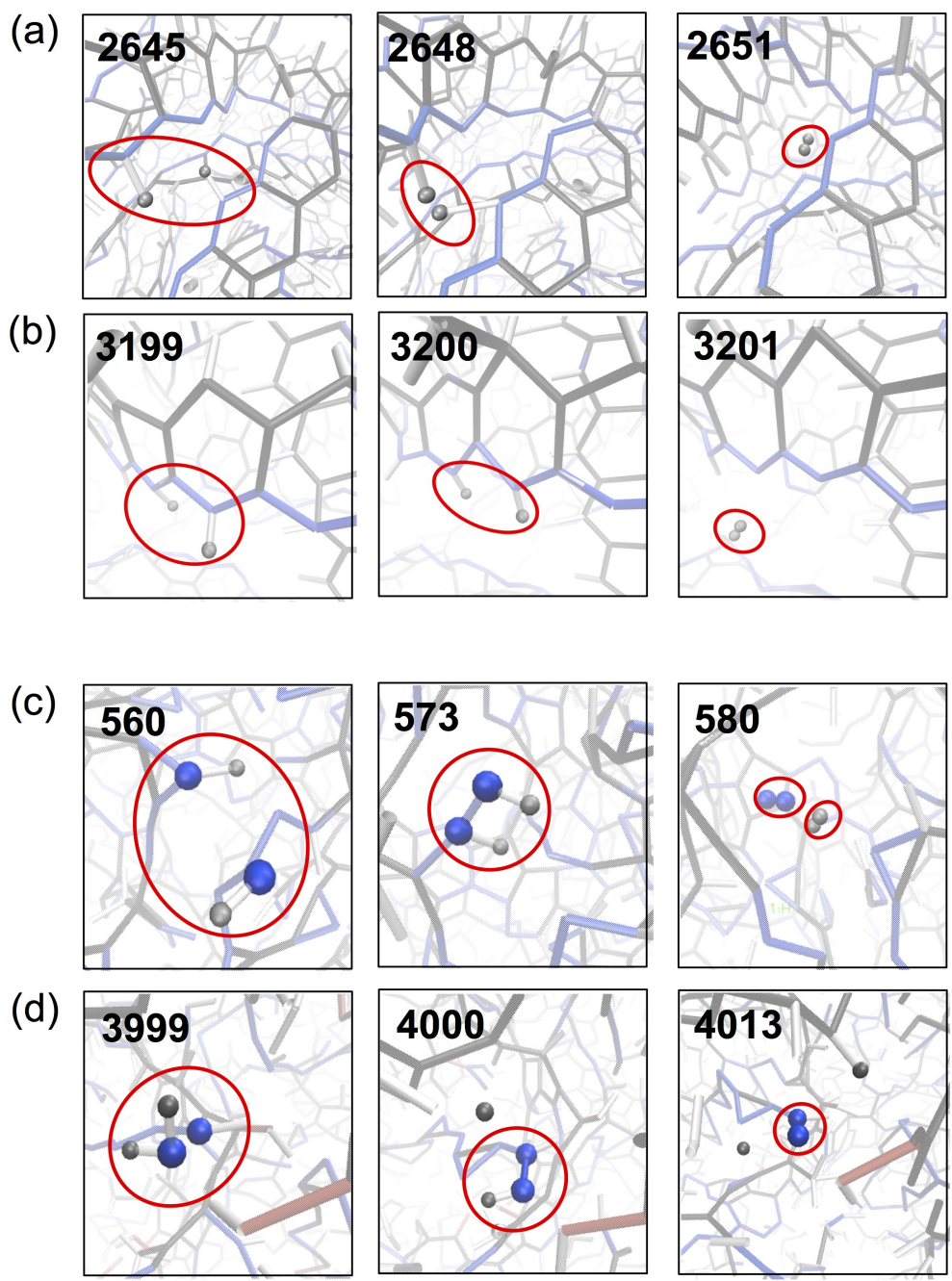

Figure 9: **The production pathways for hydrogen and nitrogen molecules**. The examples of the simulation snapshots visualizing the inter-chain (a) and intra-chain (b) hydrogen molecule production for ladder PAN. The snapshots showing the nitrogen molecule production for oxidized PAN are (c) when two –NH groups are close enough, that N-N bond can form, or (d) NH$_2$ radical and –NH are close enough that N-N bond can form. For each snapshot a frame number is indicated in the left corner. These frames were store every 6.25 fs of the carbonization process performed at NVT ensemble at T= 2800K.

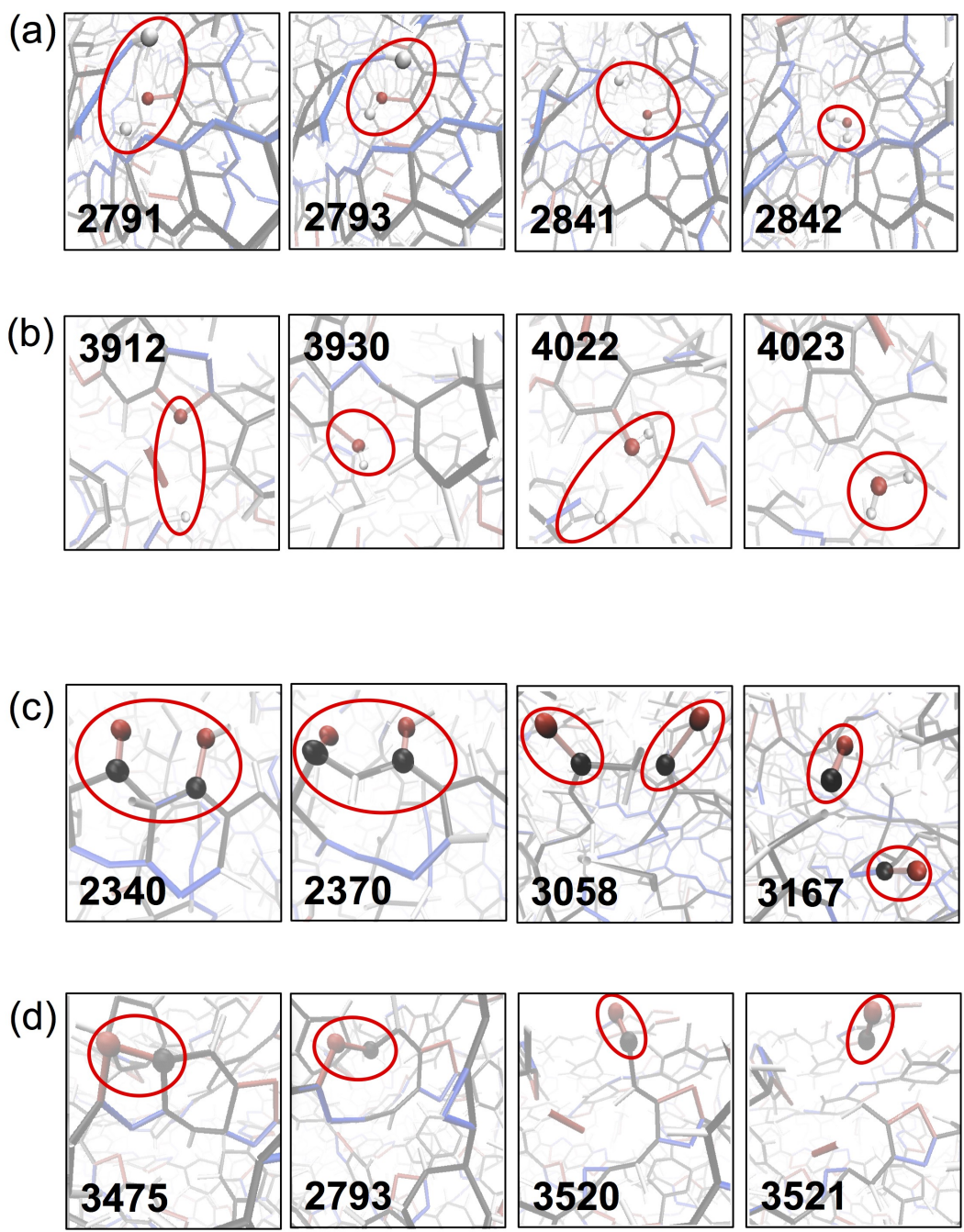

Figure 10: **The production pathways for water**, carbon monoxide and carbon dioxide molecules. The examples of the simulation snapshots visualizing: a water molecule production (a) for oxidized PAN and (b) PBO; carbon monoxide production for (c) oxidized PAN and (d) PBO. For each snapshot a frame number is indicated in the left corner. These frames were store every 6.25 fs of the carbonization process performed at NVT ensemble at T= 2800K.

The release of the gas molecules leaves mostly carbon-based, "reactive" polymers that eventually organize themselves into ring clusters. In Fig. 11 the possible pathway for evolution of such ring clusters is presented. In Fig. 11 (a-b) we see the initial 5-membered ring and then adjacent 8-membered ring creation, followed by 6-membered ring adjacent to the existing 8-membered creation on one side, and 5-membered on the other side, Fig. 11 (c-d). Next, the 7-membered and 10-membered rings are assembled alongside the existing 8- and 6-membered rings, Fig. 11 (e-f). In Fig. 11 (g) the 5-membered ring is opening and then part of this ring recombines with an existing 7-membered ring, forming a 9-membered ring, Fig. 11 (h). Then, the existing 10-membered ring is transformed into two smaller rings, 5- and 7-membered ones, Fig. 11 (i). In Fig. 11 (j) we see that on the other edge of the evolving cluster, another 5-membered ring forms. A new 6-membered ring is assembling on the edge of the existing 9-membered ring and the initial 5-membered ring, while the recently formed 5-membered ring is opening and capturing an extra carbon atom from the surroundings, Fig. 11 (k). Consequently, in Fig. 11 (l) the 9-membered ring is opening, capturing one more carbon from the surroundings and eventually closing, what leads to transformation of the 9-membered ring into a 10-membered one. The open 5-membered ring with extra carbon is also closing, constituting a new 6-membered ring. An existing connection between this new 10-membered ring and the newly formed 6-membered one is closing into another new 6-membered ring and finally the 10-membered ring is transforming into two connected 6-membered rings, Fig. 11 (m-n). This complicated pathway, that we observe in the first 100 ps of the carbonization simulations, results in the formation of all-carbon rings cluster consisting of six 6-membered rings, two 5-membered and two 7-membered rings.

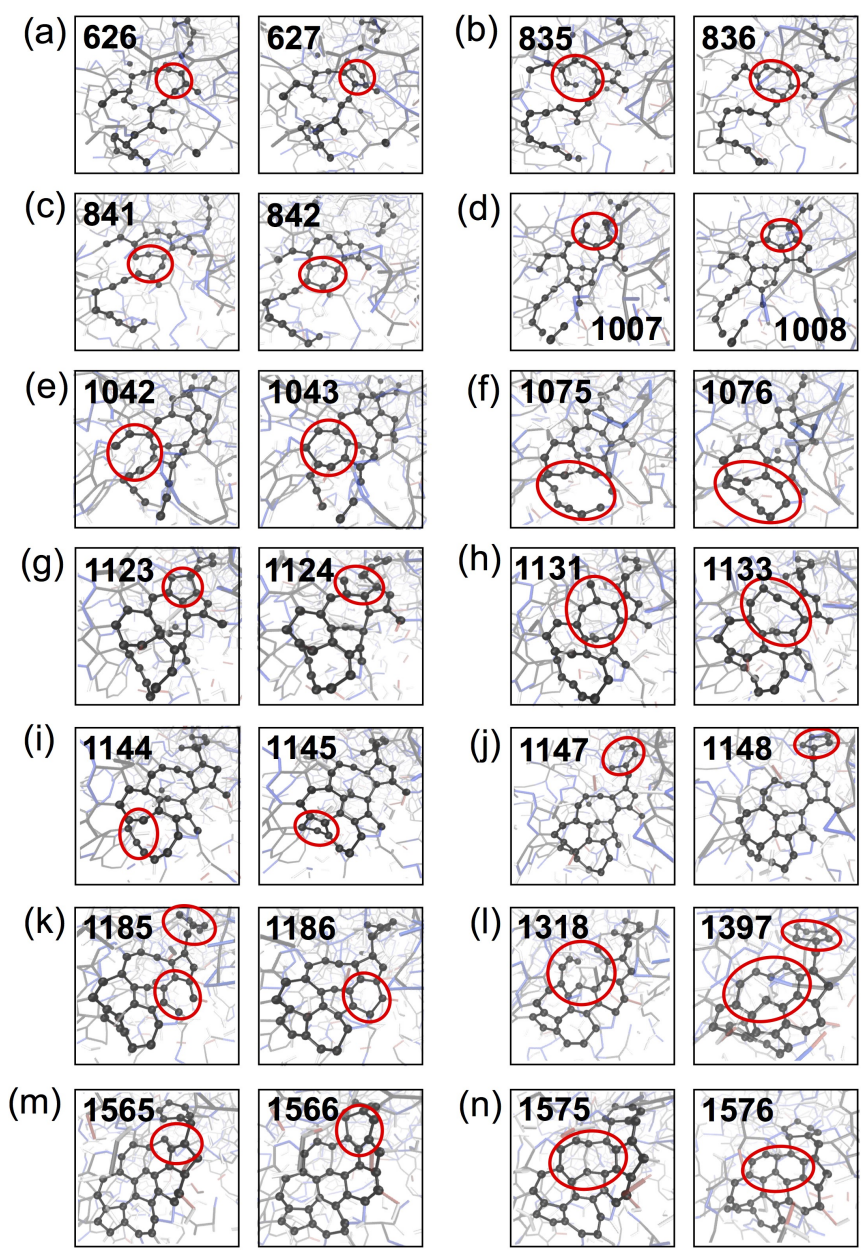

Figure 11: **All-carbon rings cluster formation**. The simulation snapshots showing the sequence of the all-carbon rings assemblies leading to the all-carbon rings formation. (a) The initial 5-membered ring and (b) the adjacent 8-membered ring creation, followed by (c) 6-membered and (d) 5-membered rings formation. Next, (e) the 7-membered and (f)10-membered rings are assembling alongside the existing 8- and 6-membered rings. (g) The 5-membered ring opening and then (h) part of this ring recombining with existing 7-membered ring, forming 9-membered ring. (i) The existing 10-membered ring is transforming into two smaller rings: 5- and 7-membered ones, and (j) another 5-membered ring forms. (k) A new 6-membered ring is assembling on the edge of the existing 9-membered ring and the recently formed 5-membered ring is opening and capturing one more carbon from surrounding. (l) The 9-membered ring opening, capturing one more carbon from surrounding and eventually closing, which leads to transformation of the 9-membered ring into 10-membered one and also already open 5-membered ring with one more carbon atom captured forms 6-membered ring. (m) Another new 6-membered ring is forming and finally (n) the 10-membered ring is transforming into two connected 6-membered rings. For each snapshot a frame number is indicated in the left corner. These snapshots were store every 6.25 fs of the carbonization process performed at NVT ensemble at T= 2800K.

An example of the final ring cluster evolved after 900 ps of the carbonization simulations at 2800K for one of the PAN samples is presented in Fig. 12 (a). As we can see, the majority of the rings are 6-membered rings, with some 5-membered, 7-membered and hetero-atoms rings. Since the 6-membered rings are the basic building blocks for the possible graphitic structure, we considered only these rings for an alignment test. The example snapshots representing the orientation of only 6-membered rings for each considered system are given in Fig. 12 (b). Qualitatively we can say that we do not observe any preferential orientation of these 6-membered clusters, which again indicates the early stage of the carbonization process. The quantitative assessment of this possible alignment can be measured by the Herman function[41], that is defined as:

$$F = \frac{1}{2}(3 < \cos^2 \varphi > - 1).$$

Where, $\varphi$ is the angle between each of rings' normal vector **e** with the resultant vector, **O** and $<>$ indicate the average value of $\cos^2 \varphi$ for all vectors in system. The normal vector **e** is calculated for each of the rings as a unit vector perpendicular to the ring plane, whereas the resultant vector **O** is given by the maximum value of the sum of the all vector in the system at the given time. For all rings perfectly aligned the value of the Herman function will be 1, if all rings have no an orientational correlation the value will be 0. The averaged values of the Herman function calculated for the last 100ps of the simulations for all considered samples are presented in Fig. 12 (c). For all cases the calculated average values are lower than 0.5, which indicates weak alignment. This gives us clear information that our 900 ps of the simulations describe only the initial stage of the carbonization process. The early stage of the carbonization is also confirmed by the composition of the final clusters calculated for all samples, Fig. 13 (a), where we see that for none of the systems does the percentage of carbon atoms exceed 90%. Nonetheless, there is a clear trend, indicating the highest carbon content for the PBO samples compared to the PAN ones. The same clear trend we observe for the 6-membered ring production, Fig. 13 (b), which indicates that PBO can be a promising candidate for the carbon fiber precursor.

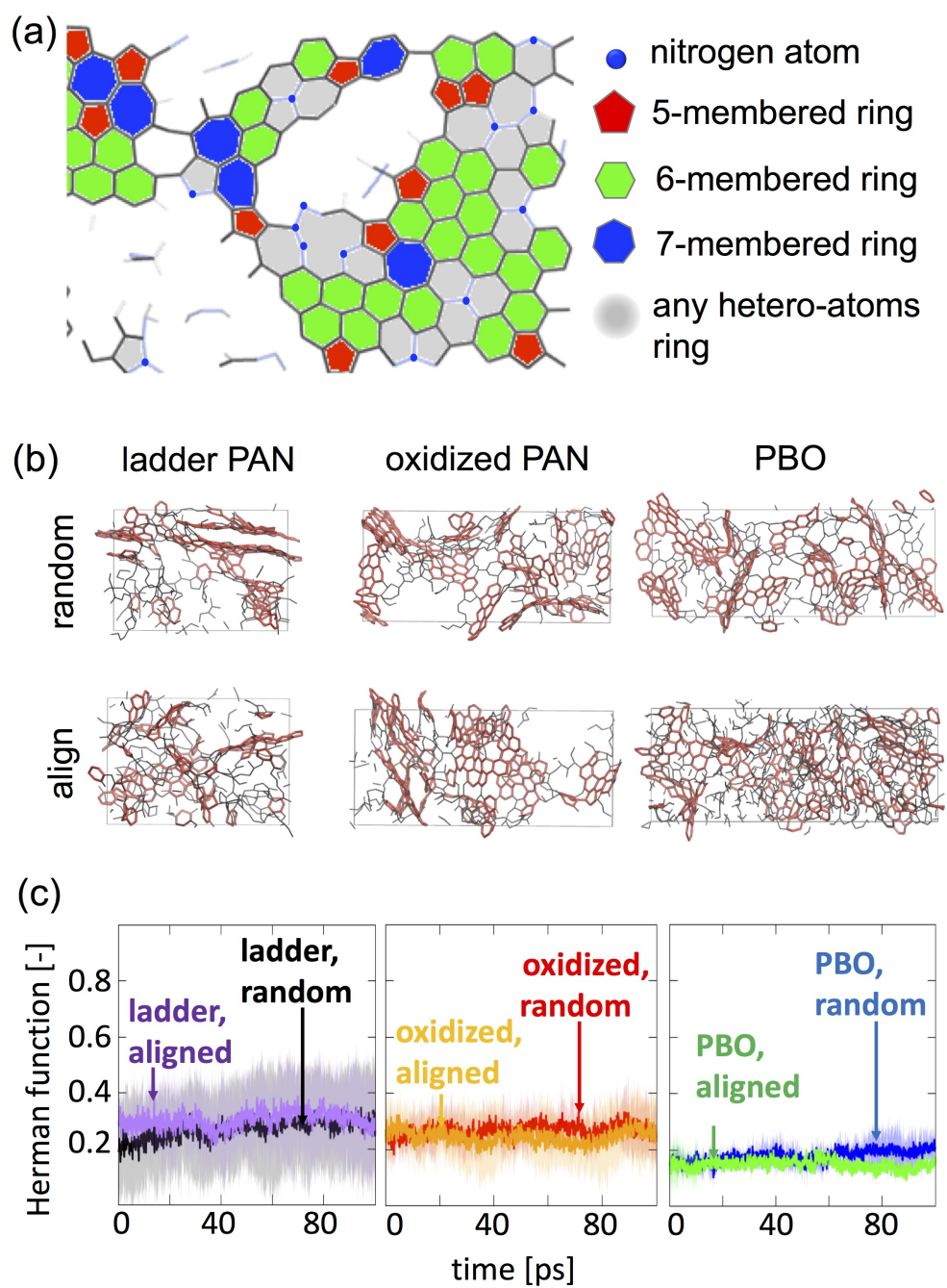

Figure 12: **The final rings clusters and 6-membered rings alignment**. (a) An example of the final rings cluster with 5/6/7-membered rings indicated as red, green, blue respectively. For clarity only bonds are represented, shown as the lines, with nitrogen atoms shown as the blue spheres. For all snapshots, (b) The examples of the final all-carbon bonds network given by solid black lines, with 6-membered rings highlighted in red for each of the considered configurations. c) The time evolution of the Herman function calculated for the last 100ps of the simulations. All curves are the average values over 3 samples and calculated errors are represented as shadow areas.

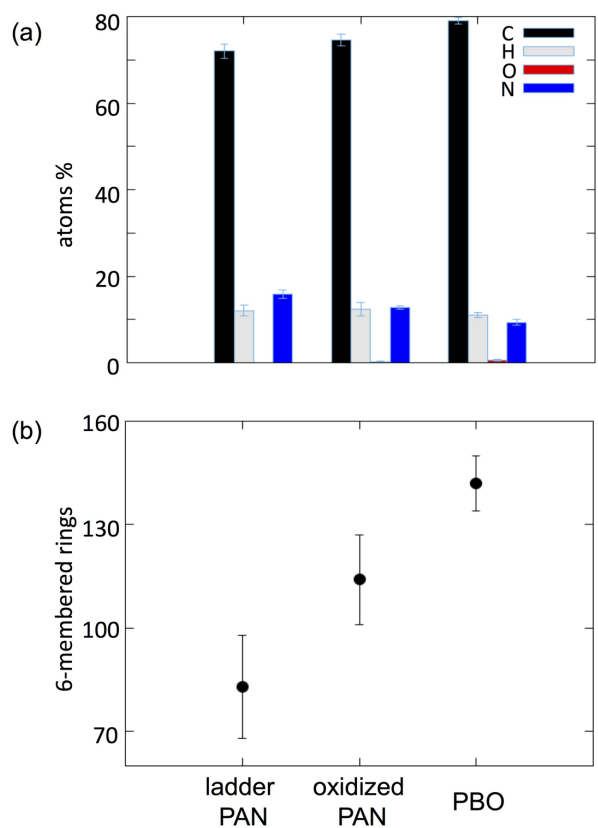

Figure 13: **The comparison of the final cluster composition and number of 6-membered rings produced.** For each considered polymer: ladder PAN, oxidized PAN and PBO the data for all 6 samples are averaged out (3 samples with polymers initially placed randomly in the box and aligned along x-direction) and standard deviations are indicated. The comparison of (a) the final cluster composition and (b) the final number of 6-membered all-carbon rings after 900ps of carbonization simulations at temperature 2800K, using NVT ensemble.

**CONCLUSIONS**

To enable simulations on the carbonization process of PAN and PBO polymers we developed a new set of ReaxFF parameters - CHON-2019 - which incorporates parameters developed by Srinivasan et al. [ref], that are suitable for reasonable graphene mechanical properties estimation and CHO-2016 parameters developed by Ashraf et al.[28], suitable for production of small molecules, like $CO_2$/CO, $H_2O$. Moreover, CHON-2019 parameters were also trained for C/N/H chemistry based on the new DFT data with a particular focus on $N_2$ formation kinetics. Equipped with this CHON-2019 ReaxFF parameters set, we were able to answer questions pertaining to the underlying molecular details of the conversion of PAN and PBO polymers and relate this to the nitrogen and other gas molecule formation, as well as to all-carbon rings production which is a starting point for evolution of graphitic structures. First, we determined that presence of the amide groups for PAN is important for effective $N_2$ production. In the case of the oxidized PAN and a possible alternative precursor, PBO, we were able reproduce the main gas products observed during the heat treatment of these polymer fibers. For oxidized PAN the main products are $N_2$, $H_2$, $H_2O$, for a closed system, and hydrogen cyanide and ammonia for system where these molecules systematically removed during the carbonization simulations. For PBO the main products we observed in our simulations are hydrogen, water and carbon monoxide. Based on the analysis of the snapshots from the carbonization simulations we were able to propose some previously unreported alternative reaction pathways for these systems.

We also analyzed the all-carbon rings production for all considered systems and were able to determine that 6-membered rings do not evolve only from 5-membered rings, but also from other all-carbon rings with more that 6 carbon atoms. Moreover, based on our analysis we can determine that PBO polymers can be a promising alternative precursor for carbon fiber productions, since we do observe the highest 6-membered rings production for any of the samples by using PBO polymers. However, based on the final cluster composition analysis and possible 6-membered rings alignment, we know that the reported carbonization simulations describe only the initial stage of the carbonization process. Longer simulations time, as well as the bigger systems, will need to be considered for a full graphitic structure to be able to evolve.

## ACKNOWLEDGEMENT

This project is funded by the Department of Energy Grant No. DE-EE0008195.

## AUTHOR CONTRIBUTION

M. K. and C. A. contributed equally to this work.

**TOC**

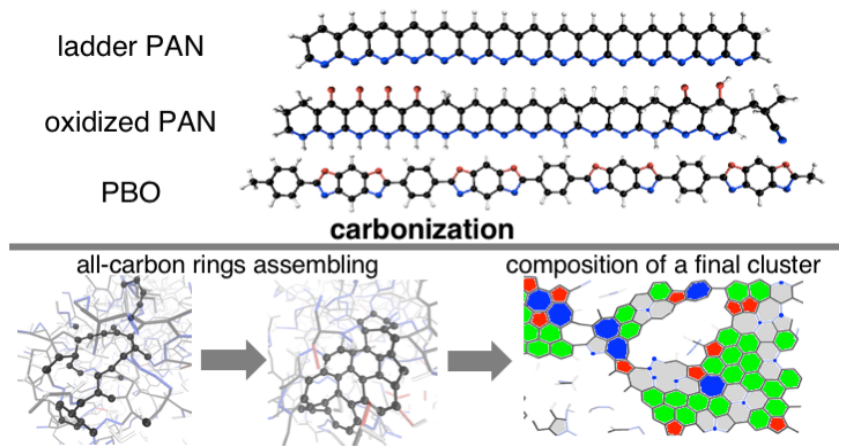

# Supporting Information

## Atomistic scale analysis of the carbonization process for C/H/O/N-based polymers with the ReaxFF reactive force field

Malgorzata Kowalik[1,‡], Chowdhury Ashraf[1,‡], Behzad Damirchi[1],

Dooman Akbarian[1], Siavash Rajabpour[2] and Adri C. T. van Duin[1,*]

[1] Department of Mechanical Engineering, The Pennsylvania State University, University Park, PA 16802, United States

[2] Department of Chemical Engineering, The Pennsylvania State University, University park, PA, 16802, United States

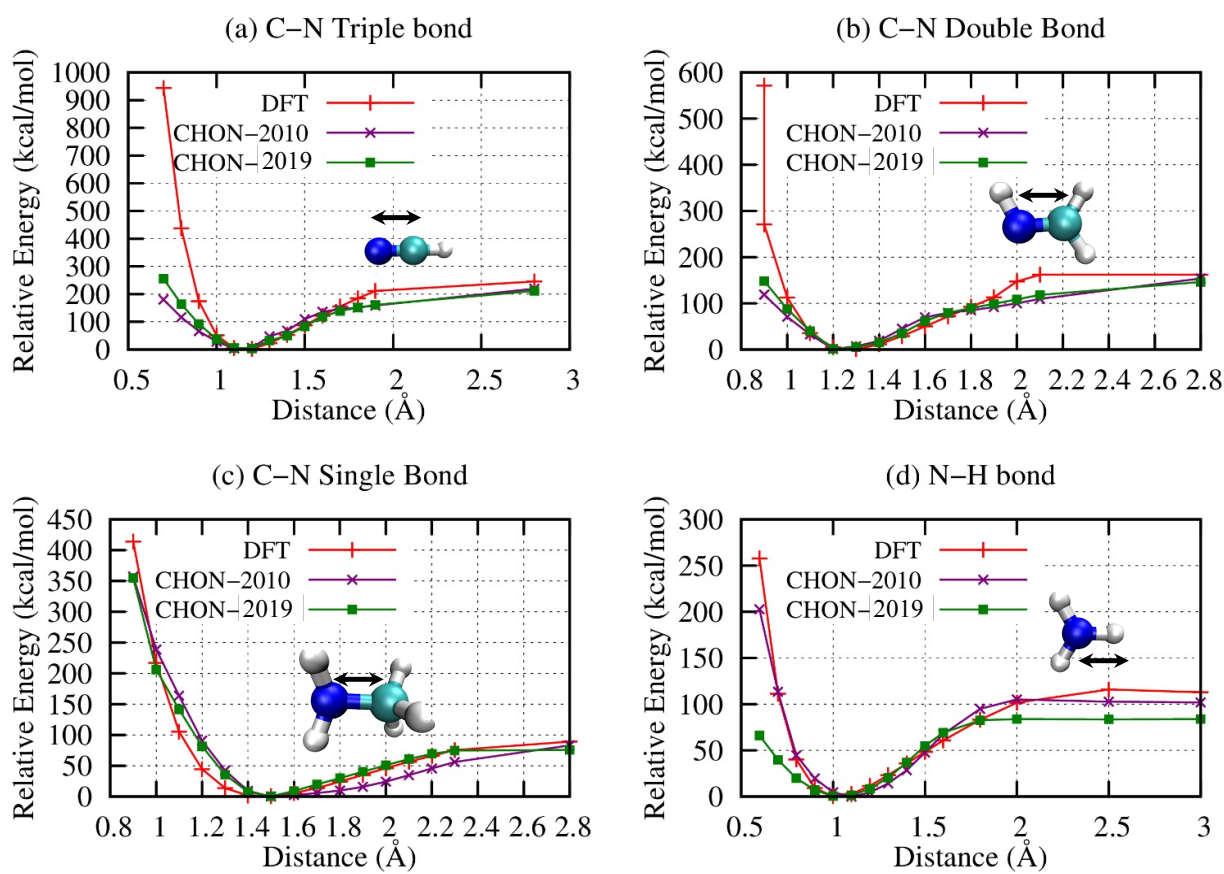

**Figure S1:** DFT(6-311G**/B3LYP) and ReaxFF (CHNO-2010 and CHNO-2019) bond dissociation energies for (a) C-N triple bond, (b) C-N double bond, (c) C-N single bond and (d) N-H bond. Cyan, red, blue and white spheres represent carbon, oxygen, nitrogen and hydrogen atom respectively

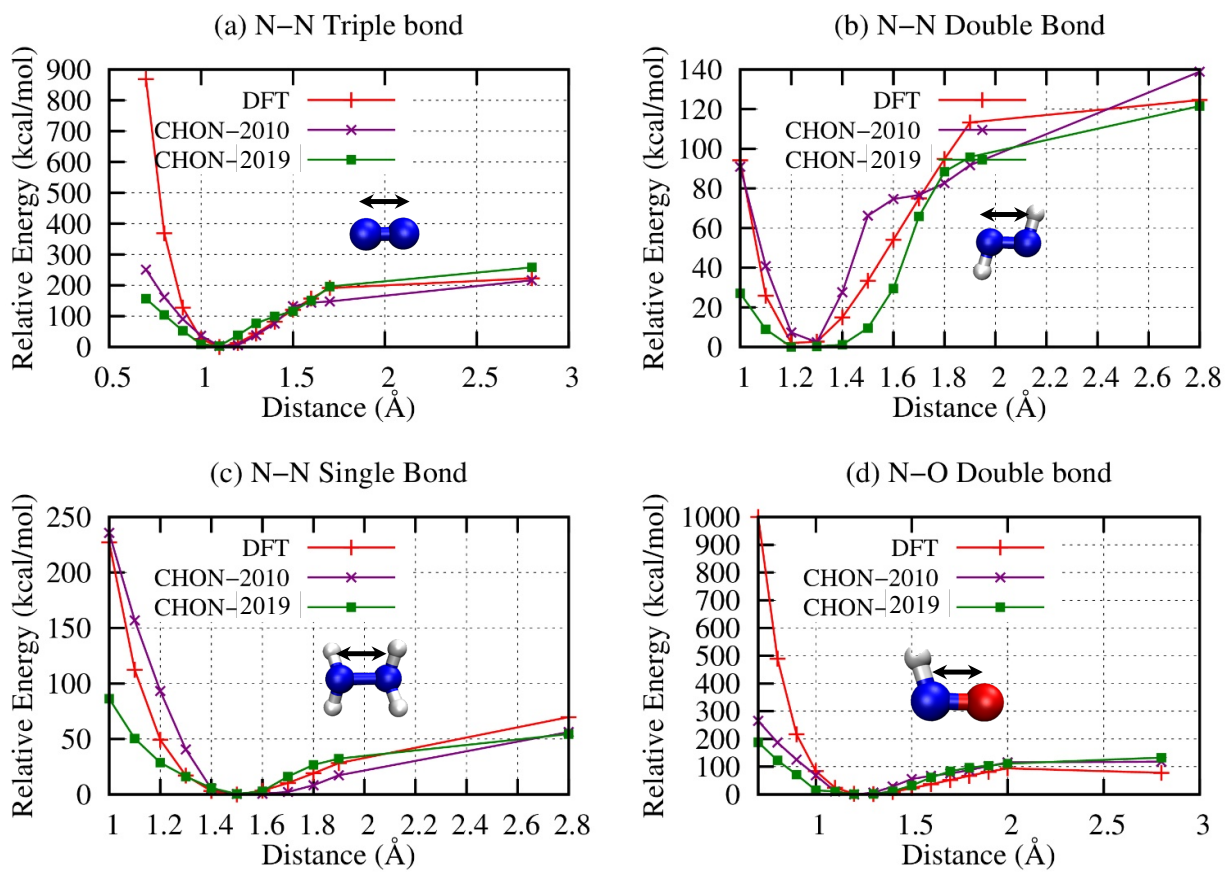

**Figure S2**: DFT(6-311G**/B3LYP) and ReaxFF (CHNO-2010 and CHNO-2019) bond dissociation energies for (a) N-N triple bond, (b) N-N double bond, (c) N-N single bond and (d) N-O double bond. Cyan, red, blue and white spheres represent carbon, oxygen, nitrogen and hydrogen atom respectively

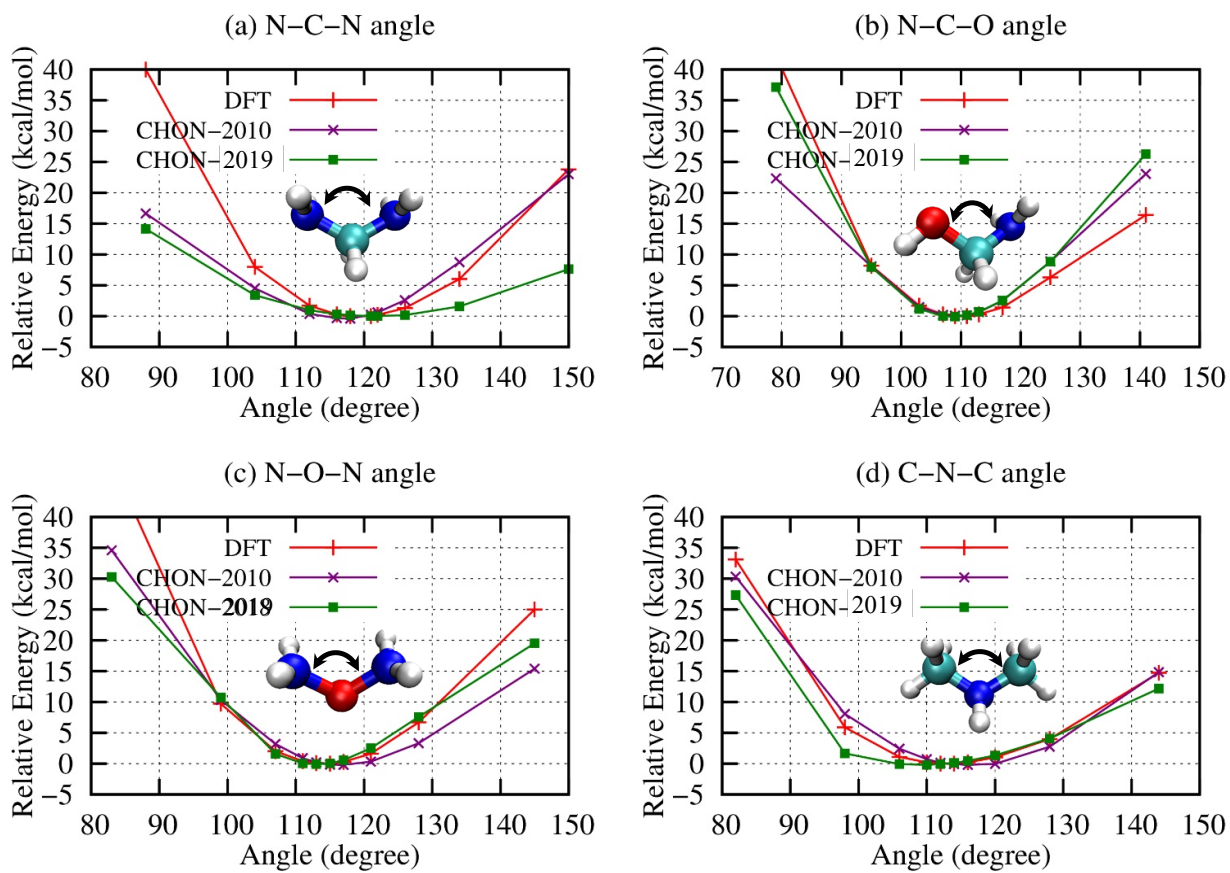

**Figure S3:** DFT(6-311G**/B3LYP) and ReaxFF (CHNO-2010 and CHNO-2019) valence angle distortion energies for (a) N-C-N angle, (b) N-C-O angle, (c) N-O-N angle and (d) C-N-C angle. Cyan, red, blue and white spheres represent carbon, oxygen, nitrogen and hydrogen atom respectively.

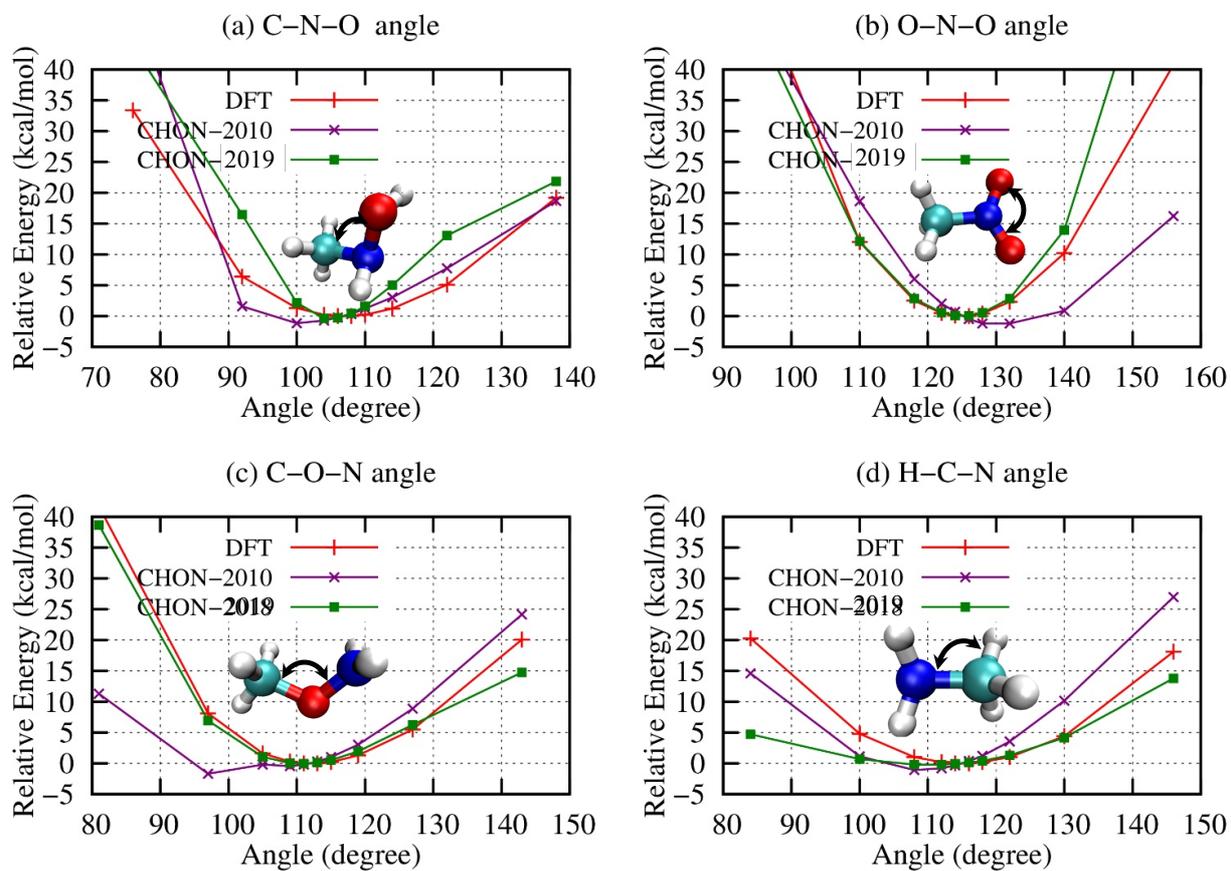

**Figure S4:** DFT(6-311G**/B3LYP) and ReaxFF (CHNO-2010 and CHNO-2019) valence angle distortion energies for (a) C-N-O angle, (b) O-C-O angle, (c) C-O-N angle and (d) H-C-N angle. Cyan, red, blue and white spheres represent carbon, oxygen, nitrogen and hydrogen atom respectively

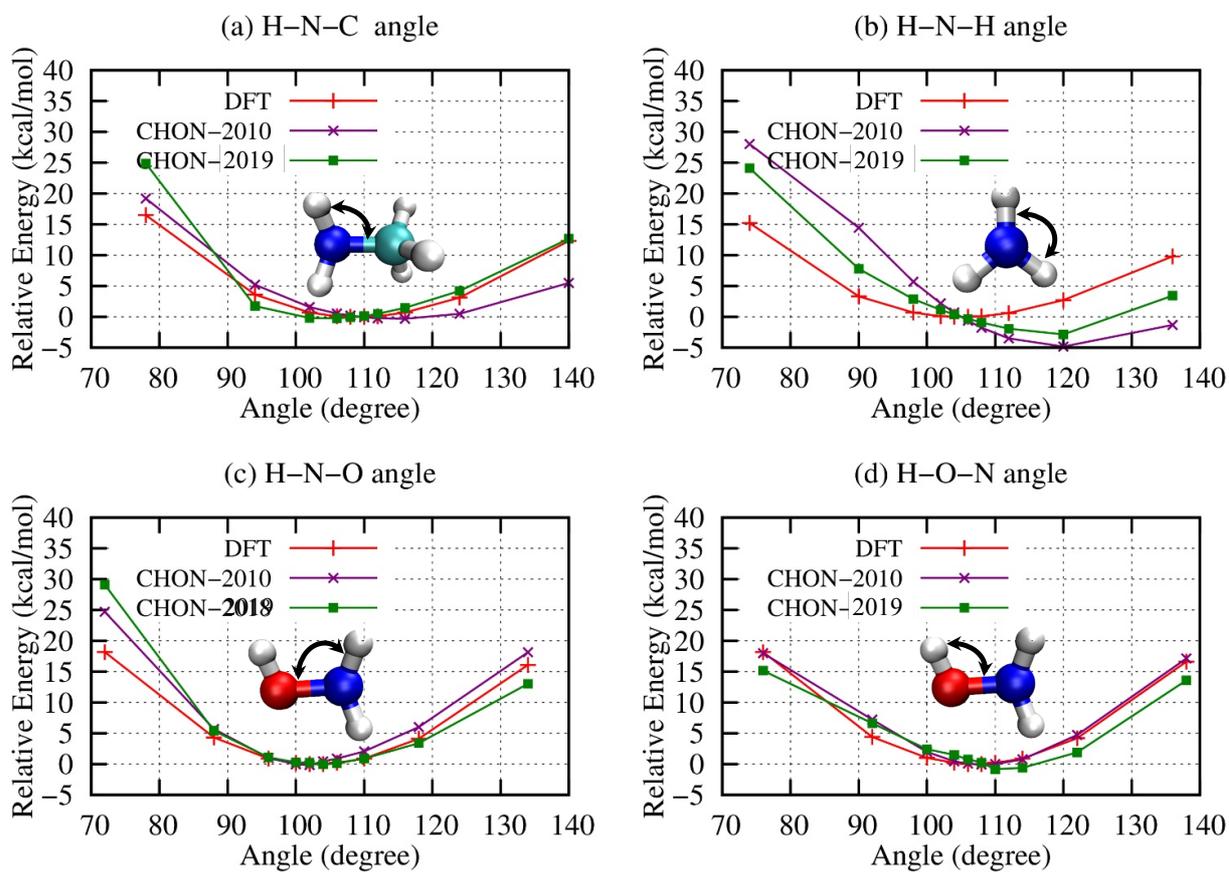

**Figure S5:** DFT(6-311G**/B3LYP) and ReaxFF (CHNO-2010 and CHNO-2019) valence angle distortion energie5 for (a) H-N-C angle, (b) H-N-H angle, (c) H-N-O angle and (d) H-O-N angle. Cyan, red, blue and white spheres represent carbon, oxygen, nitrogen and hydrogen atom respectively

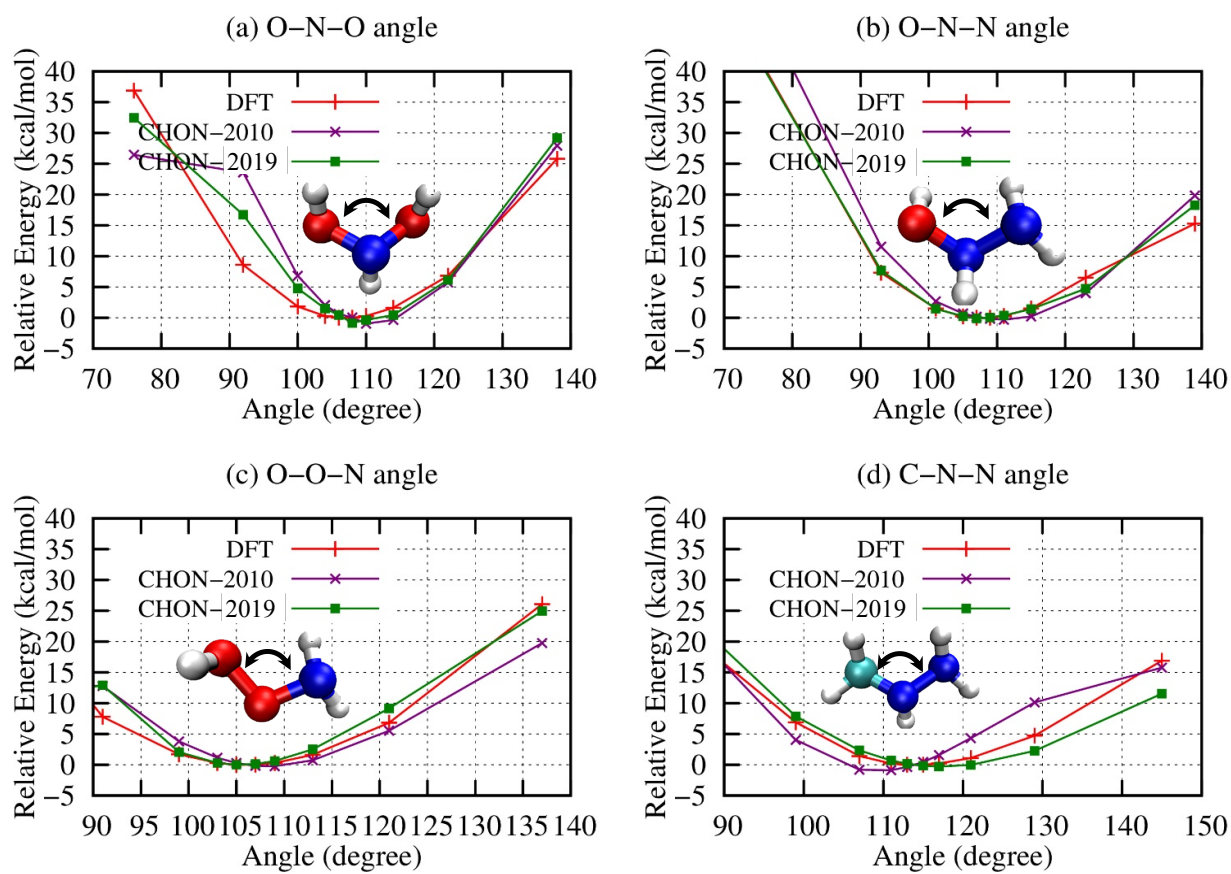

**Figure S6:** DFT(6-311G**/B3LYP) and ReaxFF (CHNO-2010 and CHNO-2019) valence angle distortion energies for (a) O-N-O angle, (b) O-N-N angle, (c) O-O-N angle and (d) C-N-N angle. Cyan, red, blue and white spheres represent carbon, oxygen, nitrogen and hydrogen atom respectively

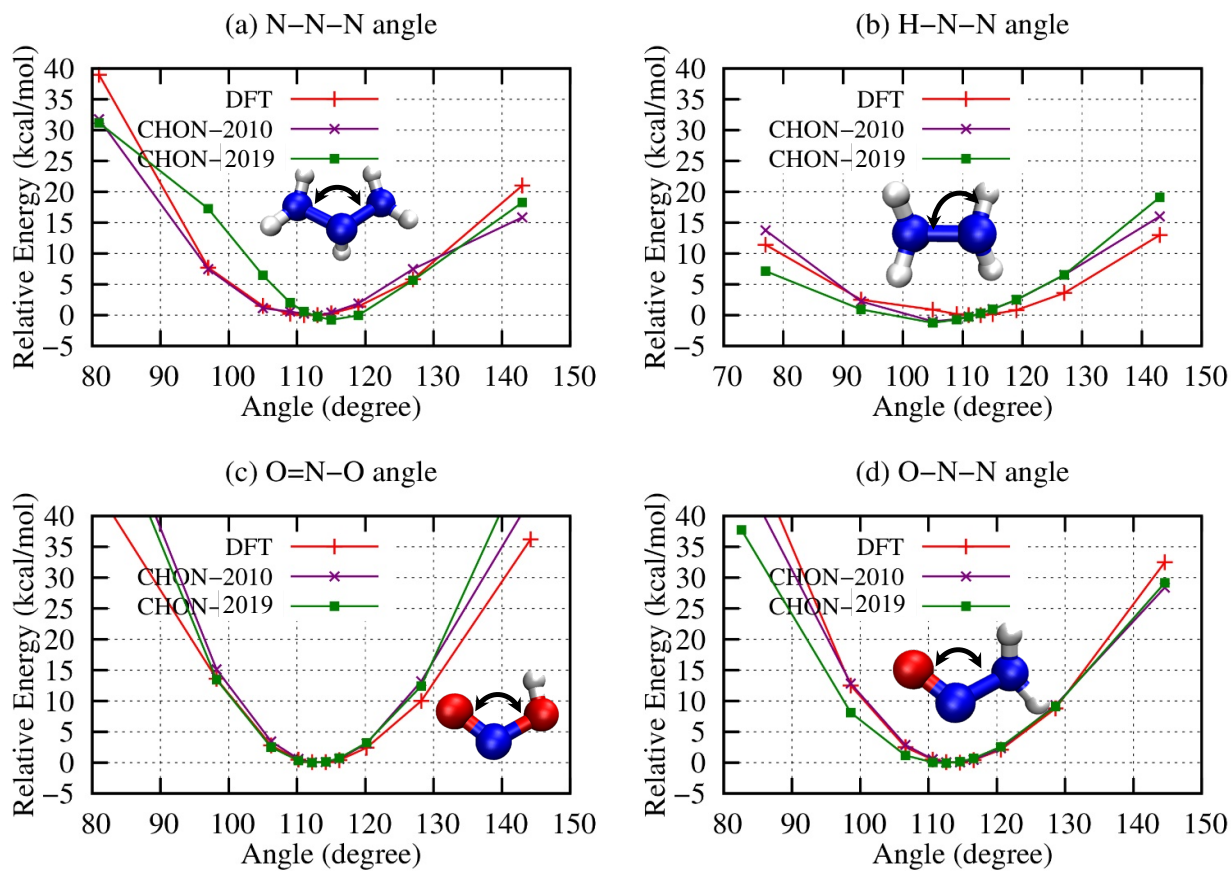

F**igure S7:** DFT(6-311G**/B3LYP) and ReaxFF (CHNO-2010 and CHNO-2019) valence angle distortion energies for (a) N-N-N angle, (b) H-N-N angle, (c) O=N-O angle and (d) O-N-N angle. Cyan, red, blue and white spheres represent carbon, oxygen, nitrogen and hydrogen atom respectively

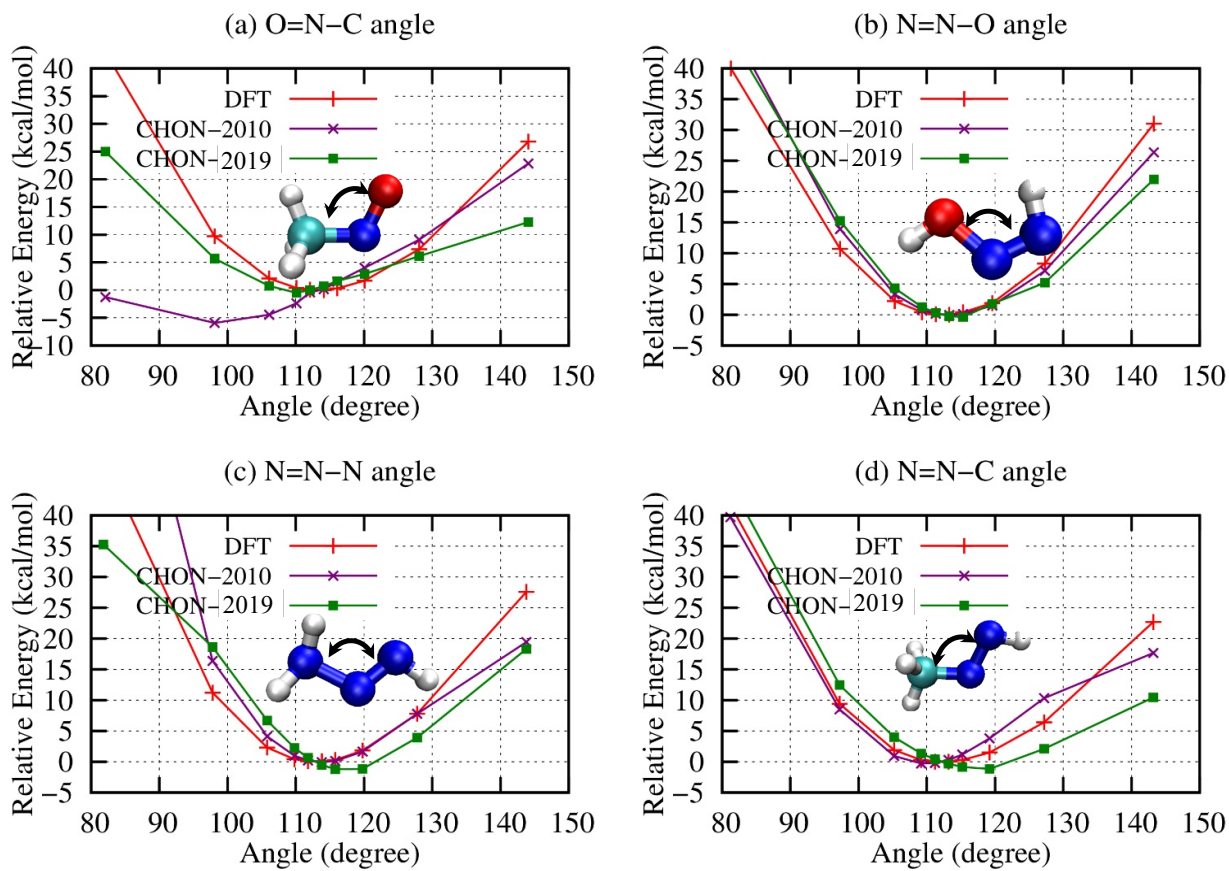

**Figure S8:** DFT(6-311G**/B3LYP) and ReaxFF (CHNO-2010 and CHNO-2019) valence angle distortion energies for (a) O=N-C angle, (b) N=N-O angle, (c) N=N-N angle and (d) N=N-C angle. Cyan, red, blue and white spheres represent carbon, oxygen, nitrogen and hydrogen atom respectively

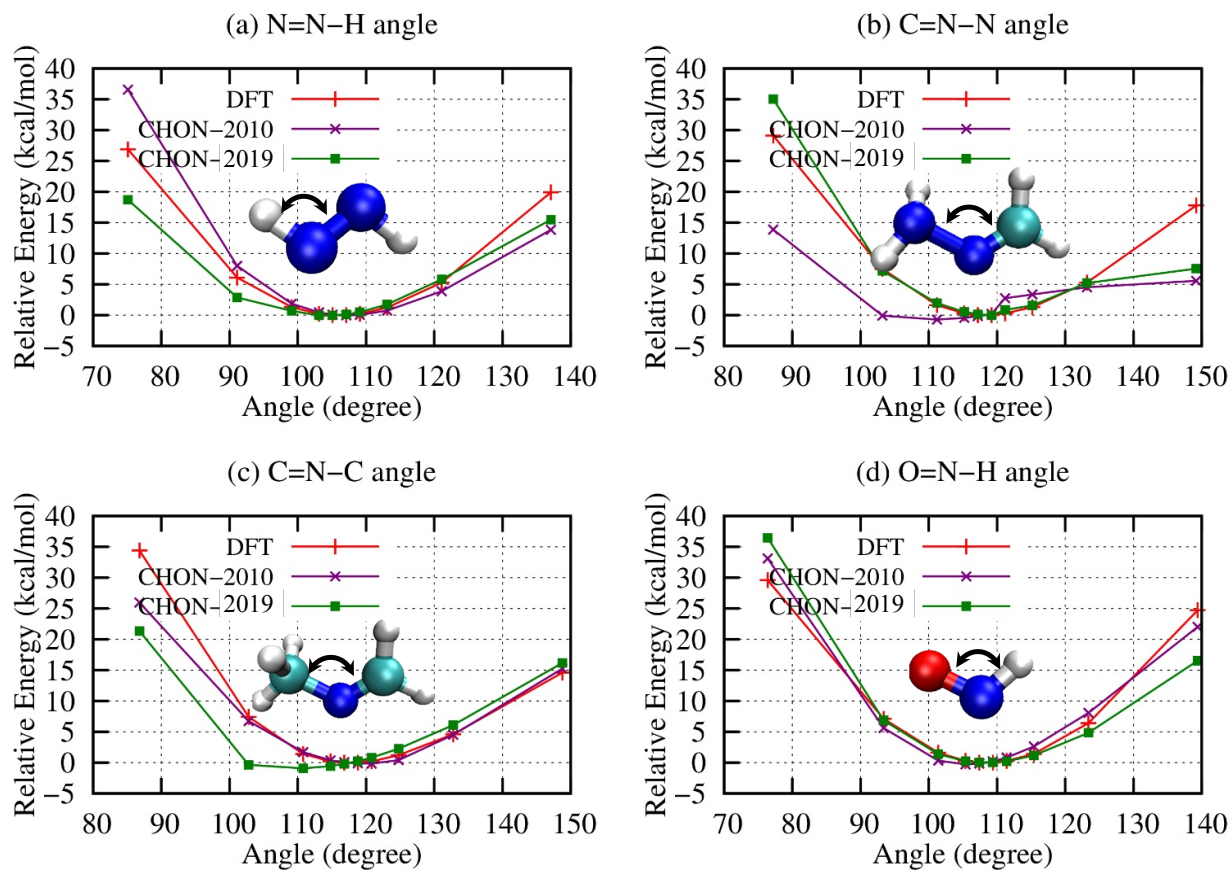

**Figure S9:** DFT(6-311G**/B3LYP) and ReaxFF (CHNO-2010 and CHNO-2019) valence angle distortion energies for (a) N=N-H angle, (b) C=N-N angle, (c) C=N-C angle and (d) O=N-H angle. Cyan, red, blue and white spheres represent carbon, oxygen, nitrogen and hydrogen atom respectively

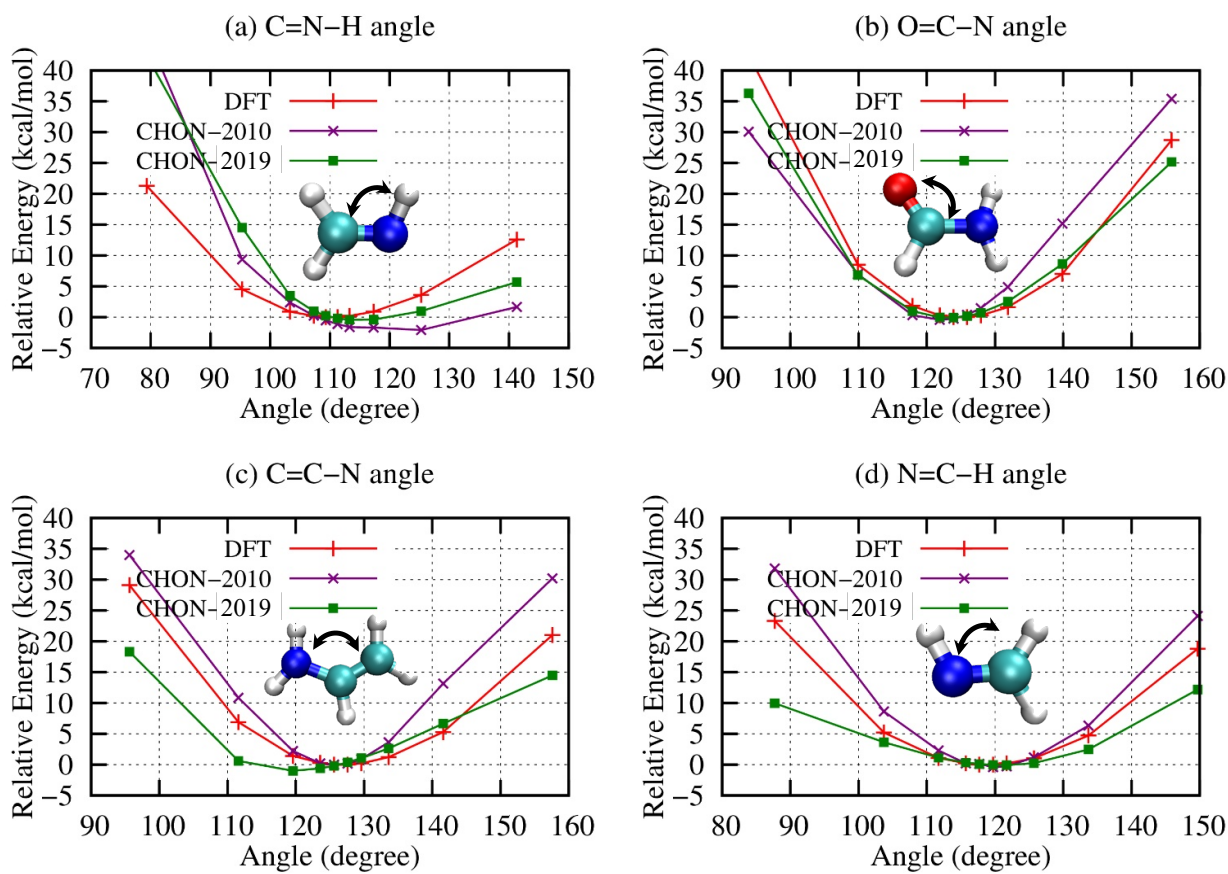

**Figure S2:** DFT(6-311G**/B3LYP) and ReaxFF (CHNO-2010 and CHNO-2019) valence angle distortion energies for (a) C=N-H angle, (b) O=C-N angle, (c) C=C-N angle and (d) N=C-H angle. Cyan, red, blue and white spheres represent carbon, oxygen, nitrogen and hydrogen atom respectively

# Supporting Information

# Atomistic scale analysis of the carbonization process for C/H/O/N-based polymers with the ReaxFF reactive force field

Malgorzata Kowalik[1,‡], Chowdhury Ashraf[1,‡], Behzad Damirchi[1],

Dooman Akbarian[1], Siavash Rajabpour[2] and Adri C. T. van Duin[1,*]

[1] Department of Mechanical Engineering, The Pennsylvania State University, University Park, PA 16802, United States

[2] Department of Chemical Engineering, The Pennsylvania State University, University park, PA, 16802, United States

## Force Field developed in this study (ReaxFF CHON-2019)

```
Reactive MD-force field: Combustion C/H/O force field + atom type N
(May11, 2018)
40       ! Number of general parameters
   50.0000 !p(boc1)
    9.5469 !p(boc2)
   26.5405 !p(coa2)
    0.6863 !p(trip4)
    2.7295 !p(trip3)
   70.0000 !kc2
    1.0588 !p(ovun6)
    4.1262 !p(trip2)
   12.1176 !p(ovun7)
   13.3056 !p(ovun8)
  -68.9784 !p(trip1)
    0.0000 !Lower Taper-radius (swa)
   10.0000 !Upper Taper-radius (swb)
    0.0000 !not used
   33.8667 !p(val7)
    6.0891 !p(lp1)
    1.0563 !p(val9)
    2.0384 !p(val10)
    6.1431 !not used
    6.9290 !p(pen2)
    0.3989 !p(pen3)
    3.9954 !p(pen4)
    0.0000 !not used
    5.7796 !p(tor2)
   10.0000 !p(tor3)
    1.9487 !p(tor4)
    0.0000 !not used
    2.1645 !p(cot2)
```

```
   1.5591 !p(vdW1)
   0.1000 !Cutoff for bond order*100 (cutoff)
   2.1365 !p(coa4)
   0.6991 !p(ovun4)
  50.0000 !p(ovun3)
   1.8512 !p(val8)
   0.0000 !not used
   0.0000 !not used
   0.0000 !not used
   1.0000 !not used
   2.6962 !p(coa3)
   2.0000 !triple bond on/off (0 for CO, 1 for CO and N2, 2 for all)
  4    ! Nr of atoms; atomID;ro(sigma); Val;atom mass;Rvdw;Dij;gamma
         alfa;gamma(w);Val(angle);p(ovun5);n.u.;chiEEM;etaEEM;n.u.
         ro(pipi);p(lp2);Heat
increment;p(boc4);p(boc3);p(boc5),n.u.;n.u.
         p(ovun2);p(val3);n.u.;Val(boc);p(val5);n.u.;n.u.;n.u.
 C   1.3674   4.0000  12.0000   2.0453   0.1444   0.9500   1.1706
4.0000
       9.0000   1.5000   4.0000  27.5134  79.5548   5.0191   7.0500
0.0000
       1.1168   0.0000     NaN  14.2732  24.4406   6.7313   0.8563
0.0000
      -4.1021   5.0000   1.0564   4.0000   2.9663   0.0000   0.0000
0.0000
 H   0.9479   1.0000   1.0080   1.1364   0.0232   0.9900  -0.1000
1.0000
       9.0643   4.7746   1.0000   0.0000 121.1250   4.7757   9.7732
1.0000
      -0.1000   0.0000     NaN   2.5194   2.3785   0.2223   1.0698
0.0000
     -15.7683   2.1488   1.0338   1.0000   2.8793   0.0000   0.0000
0.0000
 O   1.1939   2.0000  15.9990   1.9289   0.1201   0.9900   1.0981
6.0000
      10.4842   8.2916   4.0000  28.8967 116.0768   7.9703   7.0500
2.0000
       1.0479  20.0000     NaN  10.0338   2.2024   0.9942   0.9745
0.0000
      -3.6141   2.7025   1.0493   4.0000   2.9225   0.0000   0.0000
0.0000
 N   1.3638   3.0000  14.0000   1.7000   0.0967   0.8537   1.1943
5.0000
       9.8544  10.4284   4.0000  41.8891 100.0000   7.7391   7.5000
2.0000
       1.0200   0.0700     NaN   1.5271   2.9480   2.6234   0.9745
0.0000
      -5.6116   2.0047   1.0183   4.0000   2.5196   0.0000   0.0000
0.0000
 10    ! Nr of bonds; at1;at2;De(sigma);De(pi);De(pipi);p(be1);p(b
                p(be2);p(bo3);p(bo4);n.u.;p(bo1);p(bo2)
  1  1  80.8865 107.9944  52.0636   0.5218  -0.3636   1.0000  34.9876
0.7769
```

```
         6.1244  -0.1693   8.0804   1.0000  -0.0586   8.1850   1.0000
0.0000
   1  2 179.5195   0.0000   0.0000  -0.5242   0.0000   1.0000   6.0000
0.7187
         5.4740   1.0000   0.0000   1.0000  -0.1144   6.7029   0.0000
0.0000
   2  2 113.9232   0.0000   0.0000  -0.5971   0.0000   1.0000   6.0000
0.9093
         1.7152   1.0000   0.0000   1.0000  -0.0450   6.0710   0.0000
0.0000
   1  3 136.4945 164.1201   5.5000  -0.9159  -0.1075   1.0000  10.6519
0.8644
         0.6858  -0.4602   9.5754   1.0000  -0.1745   4.5987   0.0000
0.0000
   3  3 148.0798 155.2406  20.1160  -1.0000  -0.1254   1.0000  33.0027
0.7790
         0.7673  -0.1697   7.0028   1.0000  -0.1300   5.1959   1.0000
0.0000
   2  3 169.1351   0.0000   0.0000  -0.8810   0.0000   1.0000   6.0000
0.5757
         1.5482   1.0000   0.0000   1.0000  -0.1788   4.6622   0.0000
0.0000
   1  4 146.4220 161.9411  83.1445  -0.0673  -0.7385   1.0000  20.5574
0.3439
         1.1554  -0.7615   6.3243   1.0000  -0.1692   5.3062   1.0000
0.0000
   2  4 131.9942   0.0000   0.0000  -0.2031   0.0000   1.0000   4.0000
0.4507
        10.2925  -0.3653   0.0000   1.0000  -0.0527   8.0000   0.0000
0.0000
   3  4  78.7524 155.4183 100.4654   1.0000  -1.0000   1.0000  40.0000
0.1723
         0.1607  -0.5703   5.5634   1.0000  -0.1969   4.9725   1.0000
0.0000
   4  4  81.3043  99.0989 144.9704  -0.6110  -0.7864   1.0000   5.0000
0.1000
         1.0202  -0.1368   8.0395   1.0000  -0.1463   3.8325   1.0000
0.0000
   6    ! Nr of off-diagonal terms. at1;at2;Dij;RvdW;alfa;ro(sigma);r
   1  2   0.1253   1.5717   9.9736   1.2057  -1.0000  -1.0000
   2  3   0.1125   1.6311   8.7528   1.0929  -1.0000  -1.0000
   1  3   0.0953   1.7397   8.8986   1.4256   1.1067   1.1265
   1  4   0.1425   1.8737  10.3522   1.4256   1.3259   1.2082
   2  4   0.0660   1.5027   8.8662   1.0548  -1.0000  -1.0000
   3  4   0.1263   1.5263  10.0075   1.3841   1.2535   1.0000
  41    ! Nr of angles. at1;at2;at3;Thetao,o;p(val1);p(val2);p(coa1);
   1  1  1  76.1370  34.6920   1.1328   0.0000   0.0050   0.3556   1.8065
   1  1  2  68.0572   9.9461   4.7000   0.0000   0.4566   0.0000   1.8532
   2  1  2  65.6815  35.0000   1.8622   0.0000   0.0490   0.0000   1.0937
   1  2  2   0.0000   4.0000   7.2043   0.0000   0.0000   0.0000   1.0728
   1  2  1   0.0000   3.4110   7.7350   0.0000   0.0000   0.0000   1.0400
   2  2  2   0.0000  30.0000   5.6235   0.0000   0.0000   0.0000   1.0400
   1  1  3  78.3624  13.0773   9.0480   0.0000   0.1270  52.1129   2.3964
   3  1  3  76.7101  24.3833   5.8613 -21.8559   2.6395 -32.6534   3.6179
```

```
 2  1  3   79.1288   30.0000    1.4632    0.0000    0.2065    0.0000    2.0000
 1  3  1   80.7352   16.4130    4.9987    0.0000    0.0843    0.0000    1.0137
 1  3  3   85.4436   14.4937    3.9928    0.0000    1.4350   44.5320    1.1348
 3  3  3   89.9282   32.1199    2.7181    0.0000    0.3323   57.6122    1.0000
 1  3  2   82.9640   32.4874    0.8777    0.0000    0.9627    0.0000    1.0010
 2  3  3   85.7838   17.3139    1.9157    0.0000    3.6306    0.0000    2.1858
 2  3  2   84.2527   33.1226    0.6730    0.0000    0.7238    0.0000    2.4348
 1  2  3    0.0000   14.4588    3.1507    0.0000    3.4571    0.0000    1.0149
 3  2  3    0.0000    0.9696    3.6303    0.0000    0.0000    0.0000    1.6987
 2  2  3    0.0000    0.5797    1.9739    0.0000    0.0000    0.0000    2.4494
 1  1  4   89.5412   10.0000    3.1804    0.0000    0.1000    2.0000    2.0000
 3  1  4   75.6717   34.2176    2.4651    0.0000    0.2576    0.0000    1.3080
 4  1  4   56.7950   21.2889    1.1348    0.0000    0.1000    0.0000    2.0000
 2  1  4   57.5770   16.8737    1.6684    0.0000    0.9500    0.0000    1.0100
 1  2  4    0.0000    0.0100    5.6777    0.0000    0.0000    0.0000    1.2703
 1  3  4   77.6218   14.9138    3.5216    0.0000    0.6416    0.0000    1.5611
 3  3  4   87.1336   29.3985    2.1764    0.0000    0.6955    0.0000    1.8914
 4  3  4   70.2689   19.9584    4.2797    0.0000    0.6998    0.0000    1.6913
 2  3  4   73.7577   44.4943    0.5753    0.0000    3.0692    0.0000    1.5996
 1  4  1   81.0255   35.0000    0.7103    0.0000    1.6888    0.0000    1.0100
 1  4  3   90.0000   25.5053    4.4541    0.0000    2.1016    0.0000    1.2203
 1  4  4   67.3194   22.7804    1.8415    0.0000    2.0063    0.0000    1.0100
 3  4  3   74.8496   50.0000    1.6227   -6.6718    3.0000   50.0000    1.0100
 3  4  4   74.6195   47.0693    0.9622   -3.4101    2.1852    0.0000    1.7988
 4  4  4   66.6498   17.4122    7.0441    0.0000    1.1587    0.0000    1.2779
 1  4  2   90.0000   20.9302    1.2522    0.0000    0.6402    0.0000    3.0000
 2  4  3   83.0567   36.8198    1.0207    0.0000    0.8674    0.0000    3.0000
 2  4  4   79.0108   22.2517    3.0204    0.0000    0.4118    0.0000    2.9985
 2  4  2   49.0517   14.1263    7.4919    0.0000    0.1000    0.0000    1.0100
 1  2  4    0.0000    0.0019    6.0000    0.0000    0.0000    0.0000    1.0400
 3  2  4    0.0000    0.0019    6.0000    0.0000    0.0000    0.0000    1.0400
 4  2  4    0.0000    0.0019    6.0000    0.0000    0.0000    0.0000    1.0400
 2  2  4    0.0000    0.0019    6.0000    0.0000    0.0000    0.0000    1.0400
 30     ! Nr of torsions. at1;at2;at3;at4;;V1;V2;V3;p(tor1);p(cot1);n
  1  1  1  1    2.0474   32.6719    0.5282   -9.0000   -2.6449    0.0000
0.0000
  1  1  1  2    1.6328   78.4995   -0.1514   -6.9161   -2.9986    0.0000
0.0000
  2  1  1  2    2.4142   78.7025    0.3506   -8.8640   -6.9283    0.0000
0.0000
  1  1  1  3   -0.7104   22.6038    0.5309   -2.0000   -0.6614    0.0000
0.0000
  2  1  1  3    1.9323   52.9368    0.6554   -8.8118   -3.9854    0.0000
0.0000
  3  1  1  3   -1.2500    1.1248   -0.1230   -9.9453   -3.9000    0.0000
0.0000
  1  1  3  1   -0.6848   56.7751   -1.2733   -2.2937   -4.0000    0.0000
0.0000
  1  1  3  2   -1.4557   78.6279    0.9945   -3.2742   -2.4240    0.0000
0.0000
  2  1  3  1    0.6928   78.1546    0.5608   -3.1713   -3.7301    0.0000
0.0000
  2  1  3  2   -1.4343   77.0699    0.9875   -3.4139   -1.4053    0.0000
0.0000
```

```
  1  1  3  3   0.5153   2.1584   0.2000  -6.5859  -3.0000   0.0000
 0.0000
  2  1  3  3   0.2018  80.0000   0.3778  -2.5000  -2.8750   0.0000
 0.0000
  3  1  3  1  -1.9875  79.2591   1.0000  -2.4206  -3.9342   0.0000
 0.0000
  3  1  3  2  -1.1000  78.8002  -1.0000  -2.6282  -4.0000   0.0000
 0.0000
  3  1  3  3  -1.0000  83.5323   4.3660  -2.6805  -1.2938   0.0000
 0.0000
  1  3  3  1   3.4682   0.0781   0.9887  -6.1195  -0.5004   0.0000
 0.0000
  1  3  3  2   1.0000  16.5478  -1.0313  -2.0000  -2.6888   0.0000
 0.0000
  2  3  3  2   4.0818  -3.2744  -0.9664  -7.1634  -3.0000   0.0000
 0.0000
  1  3  3  3   4.2014 -10.0642   1.8690  -2.4805  -2.5000   0.0000
 0.0000
  2  3  3  3   1.0000 -10.0500  -1.0000  -2.1946  -0.5300   0.0000
 0.0000
  3  3  3  3   1.0000   1.6871   3.0000  -6.2660  -0.5500   0.0000
 0.0000
  4  1  1  4   3.0000  80.0000   2.0000  -2.0000  -1.8773   0.0000
 0.0000
  1  1  1  4   1.0676  41.9735  -0.6803  -6.3125  -3.0000   0.0000
 0.0000
  2  1  1  4   3.0000  44.9653   1.7235  -3.0352  -1.0000   0.0000
 0.0000
  0  1  4  0   1.4015  77.4788   1.0472  -6.9179  -1.7577   0.0000
 0.0000
  0  2  4  0  -3.0000   0.1000   0.0200  -2.8105   0.0000   0.0000
 0.0000
  0  3  4  0   3.0000  50.0719   0.2740  -8.0000  -1.0000   0.0000
 0.0000
  0  4  4  0   0.8759  30.0000  -1.7701  -8.0000  -1.0000   0.0000
 0.0000
  0  1  1  0   3.0000  38.1059   2.0000  -3.2272  -2.9827   0.0000
 0.0000
  4  1  4  4  -3.0000  40.0000  -1.8678  -7.3019  -1.0000   0.0000
 0.0000
  4     ! Nr of hydrogen bonds. at1;at2;at3;r(hb);p(hb1);p(hb2);p(hb3
  3  2  3   1.8130  -3.5409   2.3815  21.9463
  3  2  4   1.7753  -5.0000   3.0000   3.0000
  4  2  3   1.3884  -5.0000   3.0000   3.0000
  4  2  4   1.6953  -4.0695   3.0000   3.0000
```